\documentclass[useAMS,usenatbib]{mnras}

\usepackage{times}
\usepackage{amssymb, amsmath}
\usepackage{graphicx,rotating,amssymb,longtable,lscape}
\usepackage{epsfig}
\usepackage{enumerate}
\usepackage{hyperref}

 \usepackage{etoolbox}
 \makeatletter
 \patchcmd\@combinedblfloats{\box\@outputbox}{\unvbox\@outputbox}{}{%
   \errmessage{\noexpand\@combinedblfloats could not be patched}%
 }%
 \makeatother

\def\apj{\mbox{ApJ}}
\def\apjl{\mbox{ApJL}}
\def\apjs{\mbox{ApJS}}

\def\mnras{\mbox{MNRAS}}
\def\aj{\mbox{AJ}}
\def\araa{\mbox{ARA\&A}}

\def\nat{\mbox{Nature}}
\def\aap{\mbox{A\&A}}

\title[Dissecting the AGN in Circinus -- II]{Dissecting the active galactic nucleus in Circinus -- II. A thin dusty disc and a polar outflow on parsec scales
}

\author[Stalevski, Tristram \& Asmus]{Marko Stalevski$^{1,2}$\thanks{E-mail: marko.stalevski@gmail.com},  Konrad R. W. Tristram$^{3}$ and Daniel Asmus$^{4}$
\\
$^{1}$Astronomical Observatory, Volgina 7, 11060 Belgrade, Serbia\\
$^{2}$Sterrenkundig Observatorium, Universiteit Gent, Krijgslaan 281-S9, Gent, 9000, Belgium\\
$^{3}$European Southern Observatory, Casilla 19001, Santiago 19, Chile\\
$^{4}$Department of Physics \& Astronomy, University of Southampton, Southampton, SO17 1BJ, UK}

\begin{document}

\defcitealias{Stalevski2017}{Paper I}
\defcitealias{Vollmer2018}{V18}

\date{\today}

\pagerange{\pageref{firstpage}--\pageref{lastpage}} \pubyear{2017}

\maketitle

\label{firstpage}

\begin{abstract}
Recent observations which resolved the mid-infrared (MIR) emission of nearby active galactic nuclei (AGN), surprisingly revealed that their dust emission appears prominently extended in the polar direction, at odds with the expectations from the canonical dusty torus. This polar dust, tentatively associated with dusty winds driven by radiation pressure, is found to have a major contribution to the MIR flux from scales of a few to hundreds of parsecs. When facing a potential change of paradigm, case studies of objects with the best intrinsic resolution are essential.  One such source with a clear detection of polar dust is a nearby, well-known AGN in the Circinus galaxy. In the first paper, we successfully explained the peculiar MIR morphology of Circinus observed on large, tens of parsec scales with a model consisting of a compact dusty disc and an extended hollow dusty cone. In this work, we further refine the model on smaller, parsecs scales to test whether it can also explain the MIR interferometric data. We find that a model composed of a thin dusty disc seen almost edge-on and a polar outflow in the form of a hyperboloid shell can reproduce well the VLTI/MIDI observations at all wavelengths, baselines and position angles. In contrast, while providing a good fit to the integrated MIR spectrum, the dusty torus model fails to reproduce the spatially resolved interferometric data. We put forth the disc$+$hyperboloid wind model of Circinus AGN as a prototype for the dust structure in the AGN population with polar dust.
\end{abstract}


\section{Introduction}
\label{sec:intro}

In a widely-accepted picture, dust in active galactic nuclei (AGN) is to be found preferentially around the equatorial plane in a roughly toroidal shape. Such a structure is evoked to explain the absence of the broad emission lines in some objects and their appearance in the polarised light, and to account for the warm, mid-infrared (MIR) emission \citep{Antonucci1993, Netzer2015}. While its exact shape is unknown, this structure is usually nicknamed ``the dusty torus'' since, such a geometry, while arguably too simplified, provides the above requirements.

However, recent mid-infrared (MIR) interferometric observations resolved the parsec-scale dusty structures around a number of nearby active galactic nuclei (AGN). At odds with the classical picture, these observations revealed that thermal dust emission in AGN appears to be originating dominantly along the polar direction, following the orientation of the ionisation cone, jet, polarisation angle or the accretion disc rotational axis \citep{Honig2012, Honig2013, Tristram2014, Lopez-Gonzaga2016, Leftley2018}.
Furthermore, it was found that the polar dust emission extends on the scales of tens to hundreds of parsecs \citep*{Asmus2016}. The orientation of the polar dust emission on large and small scales roughly match, indicating that both might have a same physical origin, namely a dusty wind driven by the radiation pressure on the dust grains close to the sublimation radius \citep{Honig2012}. A number of theoretical works demonstrated that such dusty winds are indeed viable in AGN \citep[e.g.][]{Gallagher2015, Chan-Krolik2016, Chan-Krolik2017, Wada2016, Vollmer2018}. It was shown that polar winds can reproduce a standard range of AGN IR spectral features \citep{Honig2017} and explain the observed variety of their broad-band IR SEDs \citep{Lyu2018}. Together with the observational evidence that supermassive black holes are able to shape their immediate environment by radiative feedback \citep{Ricci2017}, these findings have a potential to lead us towards a new paradigm for the dust structure in AGN. 

The actual structure and physics of this polar dusty wind are still not understood. Likewise, its relation to the canonical dusty torus that was so far used to explain the spatially unresolved data with a fair amount of success \citep[see][and references therein]{RamosAlmeida-Ricci2017} is completely unclear. Is the polar dusty wind a secondary component with a minor contribution, or does it actually dominate the MIR emission? Is the torus actually only a relatively thin molecular disc that channels the material from larger scales to the vicinity of the supermassive black hole, while the polar wind is responsible for most of the obscuration and MIR emission? In order to make further progress, it is crucial to study the objects with the best intrinsic resolution, to allow us to actually resolve the dust structures. One object with prominent polar dust detection is AGN in the Circinus galaxy (hereafter Circinus), an archetypal Seyfert 2 galaxy. At the distance of $4.2\,$Mpc, AGN in Circinus is one of the best studied nearby AGN, with high spatial resolution ($1\arcsec=20\,$pc) and across a wide wavelength range and at different spatial scales. The parsec-scale MIR emission of Circinus was resolved by the MID-infrared Interferometric instrument (MIDI) at the Very Large Telescope Interferometer (VLTI) into two distinct components \citep{Tristram2007, Tristram2014}: (a) a disc-like component coinciding in size and orientation with the disc observed in maser emission \citep{Greenhill2003} and (b) a component extending in a polar direction, along the ionisation cone seen in the optical \citep{Wilson2000}. The polar component is responsible for up to $\sim80\%$ of the MIR emission on parsec scales \citep{Tristram2014}. Maser emission also reveals an outflow within $\sim1\,$pc of the nucleus, aligned with the ionisation cone \citep{Greenhill2003}. In \citet*[][hereafter Paper I]{Stalevski2017}, we presented new, high quality MIR images of Circinus obtained with the upgraded VISIR instrument at the VLT. These images feature a prominent bar (previously resolved by \citealt{Packham2005}) extending $40\,$pc on both sides of the unresolved nucleus. The bar is oriented almost perfectly along the edge of the ionisation cone, which led us to hypothesise in \citetalias{Stalevski2017} that it represents an edge-brightened dusty wall enveloping or coinciding with the outer boundary of the cone. The dusty cone wall is illuminated by the anisotropic accretion disc tilted towards the MIR-bright edge of the cone. Thus, the opposite edge of the dusty cone wall remains invisible. Such a tilt of the accretion disc is supported by the orientation of the inner part of the warped maser disc and the anisotropic illumination pattern seen in [\ion{O}{iii}] and H$_{\alpha}$ \citepalias[for a schematic and more details see][]{Stalevski2017}. We employed a radiative transfer code to simulate the dust emission in Circinus with the combination of a compact dusty disc and a large-scale hollow dusty cone. From the resulting model images, we produced synthetic MIR observations, corresponding to the images obtained with VISIR. We showed that this model successfully reproduces the observed MIR morphology of the AGN in Circinus, as well as its entire IR spectral energy distribution (SED), and resolved photometry extracted from small apertures at different positions along the polar bar and perpendicular to it. We invite the reader to examine \citetalias{Stalevski2017} and references therein for a deeper overview of the relevant Circinus observations, a detailed presentation of the model, as well as for further discussion on the properties of polar dust in AGN, including its origin.

In this work we proceed with the modeling of the interferometric VLTI/MIDI data. Our aim is to constrain the parsec-scale dust structure in Circinus and see if it is consistent with our large-scale model from \citetalias{Stalevski2017}, or with a canonical torus. In \S \ref{sec:mod} we present the geometry, dust properties and other relevant details of the considered models. The results of the radiative transfer simulations, the comparison with the observations and related discussions are given in \S \ref{sec:res}. Summary and conclusions are laid out in \S \ref{sec:sum}.

\section{Observational data}
\label{sec:obs}

VLTI/MIDI \citep{Leinert2003} combines the light of two telescopes to produce spectrally dispersed interferograms between 8 and 13$\,\micron$. The interferometric measurements of this instrument are the correlated flux spectra $F_{\text{cor}}$ and the differential phases, as well as the masked total flux spectra $F_{\text{tot}}$ from the single-dish measurements. The correlated fluxes  and the differential phases depend on the projected separation and orientation of the pair of the telescopes used in the observation, that is, on the projected baseline length, BL, and its position angle, PA. Each measurement represents a point of the Fourier transform of the source surface brightness distribution in the so-called \emph{uv} plane. The short baselines are sensitive to the larger scale structures, while the long baselines are probing the smaller scales. With good coverage of the \emph{uv} plane, i.e. with a sufficient number of the different combinations of the projected baselines and their position angles, the observations contain enough information to constrain the morphology of the source emission. The correlated fluxes encode how much a source is resolved at a given position angle: high $F_{\text{cor}}$ means that the source is essentially unresolved, while low $F_{\text{cor}}$ value indicates that the source is well resolved, i.e. its surface brightness is extended along this direction.

Circinus was observed in 18 epochs between 2004 and 2011 with the MIDI long baselines pairs of the 8.2 m Unit Telescopes (UTs) and with the short baselines pairs of the 1.8 m Auxiliary Telescopes (ATs), leading to a total of 152 correlated flux spectra and differential phases between 8 and 13$\,\micron$. A comprehensive presentation of the observations, the data reduction and the relevant analysis can be found in \citet{Tristram2007, Tristram2014}.

\citet{Tristram2014} modeled the MIDI data of Circinus using black body emitters with a Gaussian brightness distribution. They decomposed the parsec-scale dust distribution into a disc-like component ($\sim0.2\times1.1\,$pc) coinciding in size and orientation with the disc observed in maser emission \citep{Greenhill2003} and a component extending in a polar direction ($\sim0.8\times1.9\,$pc), along the ionisation cone seen in the optical \citep{Wilson2000}. The polar component is responsible for up to $\sim80\%$ of the MIR emission on parsec scales; the quoted sizes correspond to the full-width-half-maximum ($\mathrm{FWHM}$) of the mentioned black body Gaussians.

\section{Models of the parsec-scale dust structure in the Circinus AGN}
\label{sec:mod}

\subsection{Geometry}
\label{sec:geo}

\subsubsection{disc+polar dust}
\label{sec:pol}

In \citetalias{Stalevski2017}, we found that the ten to hundred parsec scale MIR morphology seen in VISIR images can be reproduced by either of two geometries for the dust in the polar region: a conical shell or a one-sheeted hyperboloid shell. The two geometries are very similar, being effectively the same beyond a couple of parsecs, and different only in the innermost region. Here, the inner and outer cone walls continue to converge towards the centre, while the hyperboloid walls rise almost vertically from the equatorial plane and then bend to asymptotically approach the boundary of the cone surfaces. Since the inner region is unresolved in the VISIR images ($<0.4\arcsec$), both geometries produced effectively the same results when convolved with the corresponding point spread function to produce the synthetic observations. However, it is exactly this spatial scale that is resolved with the interferometric MIDI observations. 

Guided by the quoted sizes of the disc-like and the polar components from \citet{Tristram2014} and insights from the modeling of the large-scale structure in \citetalias{Stalevski2017}, we set the model geometry to a geometrically thin disc and a hyperboloid shell as depicted in the schematic in Fig.~\ref{fig:modrgb}. Here, the parameters of the disc are: the disc flaring angle, $\Delta_{\text{dsk}}$ (measured from the equatorial plane to the edge of the disc), the outer radius of the disc, $R_{\text{out}}$, the exponent of the radial profile of the dust density, $p$ in $\rho\left(r \right)\propto r^{-p}$, and the edge-on optical depth at $9.7\,\micron$, $\tau^{dsk}_{9.7}$, setting the total dust mass in the disc. The inner radius is determined by the distance at which the dust sublimates for a given AGN luminosity. The hyperboloid shell is defined by the position of its inner and outer walls in the equatorial plane, ($a_{\text{in}}$, $a_{\text{out}}$), its angular width, $\Theta$ (the angular distance between the inner and outer cone walls towards which hyperboloids asymptotically approach), a half-opening angle, $\Delta_{\text{hyp}}$ (measured from the polar axis to the inner wall), and the total dust mass defined by an edge-on optical depth, $\tau^{hyp}_{9.7}$.

To determine how well the MIDI data can constrain the parsec-scale structure, we also tested disc$+$cone geometry. The parameters that describe a cone geometry are the same as for a hyperboloid, apart from $a_{\text{in}}$ and $a_{\text{out}}$, which are not defined.

\subsubsection{Dusty torus}
\label{sec:tor}

In \citetalias{Stalevski2017}, we also tested if the large-scale (tens of parsecs) MIR morphology can be reproduced by a  dusty torus surrounded by diffuse, ambient dust of the host galaxy in a form of a spherical shell. The torus opening angle could collimate the radiation from the accretion disc, which would illuminate diffuse dust only along the polar direction and thus, potentially result in a bar-like feature in MIR. Our investigation ruled out this case since this dust configuration could not reproduce the observed morphology. However, while the dusty torus is not responsible for the large-scale morphology, it might reproduce the interferometric data that are probing the small, parsec-scales. Viewed under a favourable angle, i.e., slightly tilted towards the observer, the torus could hide the central engine and at the same time partially expose warm dust emission from its inner surface, effectively appearing as extended in the polar direction, as suggested by \citet{Tristram2014} for Circinus and by \citet{Lopez-Rodriguez2018} in the case of NGC 1068. 

The AGN dust emission models in the literature \citep{Fritz2006, Nenkova2008, Honig-Kishimoto2010, Stalevski2012a, Siebenmorgen2015, Stalevski2016} assume a flared disc geometry, but refer to it as ``the dusty torus'', since this is a commonly accepted nickname for the parsec-scale, warm, dusty obscuring structure required by the AGN unification scheme. We used the \textsc{SKIRTOR}\footnote{\url{https://sites.google.com/site/skirtorus/}} library of the clumpy two-phase dusty torus emission models \citep{Stalevski2012a, Stalevski2016}, which was calculated with the same radiative transfer code (\S \ref{sec:RT}) that we used for the present work. This library was previously successfully used in a number of works to model and interpret the IR emission of AGN \cite[e.g.][]{Lira2013, Duras2017, Ohyama2017}. The ``dusty torus'' of SKIRTOR library consists of a large number of high-density clumps submerged into a low-density dust bounded by a flared disc geometry \cite[see Fig. 1 in][]{Stalevski2016}.
Currently, the \textsc{SKIRTOR} library contains 19200 SEDs computed for different viewing angles and a range of values of the parameters defining the torus, such as, the amount of the dust ($\tau^{dsk}_{9.7}$), the half-opening angle ($\Delta_{\text{tor}}$, measured from the polar axis to the edge of the flared disc), the ratio of outer to inner radius ($R_{\text{out}}/R_{\text{in}}$), the density gradients in the radial and polar direction (defined by $p$ and $q$ in the $\rho\left(r,\theta \right)\propto r^{-p}e^{-q|\cos\theta|}$ density law). The parameters of the clumps and inter-clump medium are fixed to the values that result in a two-phase medium with the volume filling factor $\sim0.25$ and contrast between the high- and low-density phases of $\sim100$. Further details of the model can be found in \citet{Stalevski2012a, Stalevski2016}.

\subsection{Dust properties}
\label{sec:dust}

For the polar dust models, the dust composition and grain size distribution is the same as in \citetalias{Stalevski2017}: a mixture of silicates ($53\%$) and graphite ($47\%$) in the disc and only graphite in the polar region, in both cases following the MRN power-law size distribution ($\propto a^{-3.5}$) \citep*{MRN1977} with grain sizes in the range of $a=0.1-1\,\micron$. The motivation for the chosen dust properties is detailed in Section~3.3 of \citetalias{Stalevski2017}; in short, a number of observational pieces of evidence, such as flat extinction curves and silicate feature profiles, favour grain sizes larger than in the standard Galactic interstellar medium \citep{Gaskell2004, Shao2017, Xie2017}. As for the dust in the polar region, it should reflect the dust composition at its origin, i.e. a very hot inner region where only graphite grains can survive \citep{Draine1984, Draine-Lee1984, Aitken1985, Barvainis1987, BaskinLaor2018}. The optical properties of the dust grains are taken from \citet{Laor-Draine1993} and \citet{Li-Draine2001}.

The dusty torus models were calculated with dust properties usually assumed by similar models in the literature \citep[e.g][]{Fritz2006, Nenkova2008}: a 53-47 mixture of silicates  and graphite and the MRN power-law size distribution, but with grain sizes in the range of $a=0.005-0.25\,\micron$. 
A different choice of the dust mixture would not qualitatively affect our results.

A foreground absorbing screen is applied to all the models to account for extinction by dust in the disc of the host galaxy. We constructed a simple extinction screen based on the work of \citet{Roche2006}, who employed high spatial resolution spectroscopy with T-ReCS on Gemini-South to measure depths of the $9.7\,\micron$ silicate absorption feature extracted from the different positions along the polar bar and perpendicular to it. The extinction function of this screen is taken from \citet{Tristram2007} and corresponds to a dust mixture with the same MRN power-law size distribution and grain size range of $a=0.005-0.25\,\micron$, but with a modified $9.7\,\micron$ silicate feature profile. The adopted silicate feature profile is the same as found by \citet{Kemper2004} towards the Galactic center, which we take as the best guess of the extinction towards the nucleus of the Circinus galaxy. For more details on the extinction screen see Section~4.2 in \citetalias{Stalevski2017}.

Dust properties of the different models and their components are summarised in Table~\ref{tab:dust}.

\begin{table}
\centering
\caption{Dust properties of the different models and their components.}
\label{tab:dust}
\begin{tabular}{llcc}
                                                               \hline\hline
model/component	& composition		& grain size [$\micron$]	& $\tau_{\text{V}}/\tau_{9.7}$ \\ \hline
disc		  & graphite$+$silicate	& 0.1--1			& 9.4	\\
hyperboloid	  & graphite		& 0.1--1			& 21.4	\\
torus		  & graphite$+$silicate	& 0.005--0.25			& 18.8	\\
foreground screen & graphite$+$silicate	& 0.005--0.25			& 12 \\ \hline
\end{tabular}

{\it -- Notes:} 
See \S \ref{sec:dust} for more details.
\end{table}

\subsection{Radiative transfer simulations}
\label{sec:RT}

To calculate how the above described model would appear in the IR, we employed \textsc{SKIRT}\footnote{\url{http://www.skirt.ugent.be}}, a state--of--the--art 3D radiative transfer code based on the Monte Carlo technique \citep{Baes2011, Baes-Camps2015, Camps-Baes2015}. The primary source of the photons is an accretion disc approximated by a point-like source with anisotropic emission pattern to account for the change in the projected surface area and for the limb darkening effect \citep{Netzer1987}, while its SED is described by a standard composition of power-laws \citep{Stalevski2016}. The accretion disc luminosity is allowed to vary within the range of $L_{\text{AGN}}=6\times10^{9}-7\times10^{10} L_{\odot}$, as inferred from X-ray \citep{Arevalo2014, Ricci2015}, IR \citep{Tristram2014} and optical \citep{Oliva1999} observations. The size of the simulation box is $14\times14\,$pc. 
The dust is distributed on an adaptive hierarchical octree grid with a large number of cells whose size is set by an algorithm ensuring the proper sampling of the dust density and optimal usage of the computational memory \citep{Saftly2013}. In the case of our model, the grid typically consists of $\sim7\times10^{6}$ cells of $\sim0.01\,$pc in size.
The photons are propagated through the simulation box following the standard Monte Carlo radiation transfer prescriptions that take into account all the relevant physical processes: anisotropic scattering, absorption and thermal re-emission. At the end of the simulation, images and SEDs of the model can be reconstructed for the comparison with the observations in the entire wavelength range of interest.

\begin{figure*}
\centering
\includegraphics[width=0.49\textwidth]{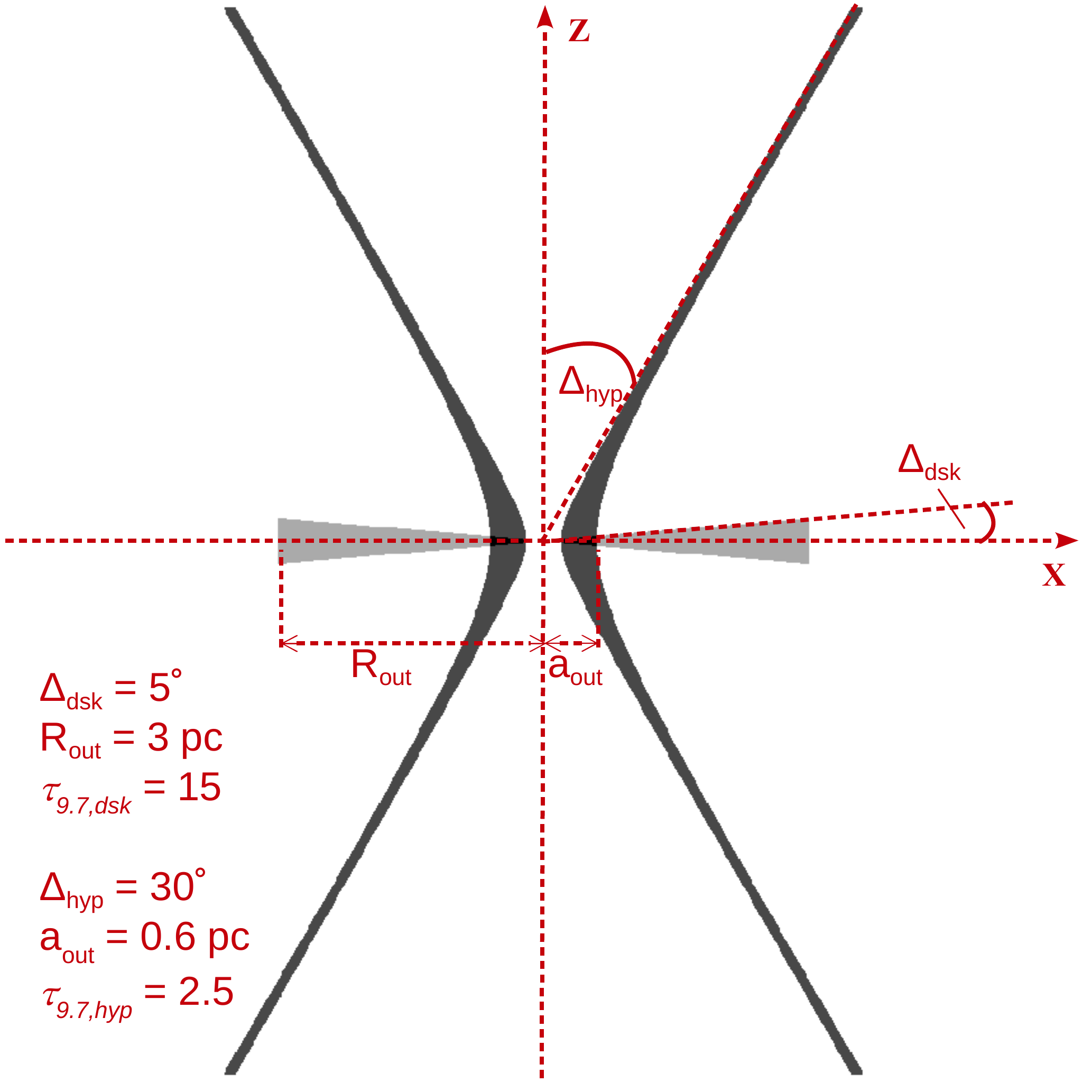}
\includegraphics[trim={0cm 0cm 0cm 0cm},clip,width=0.49\textwidth]{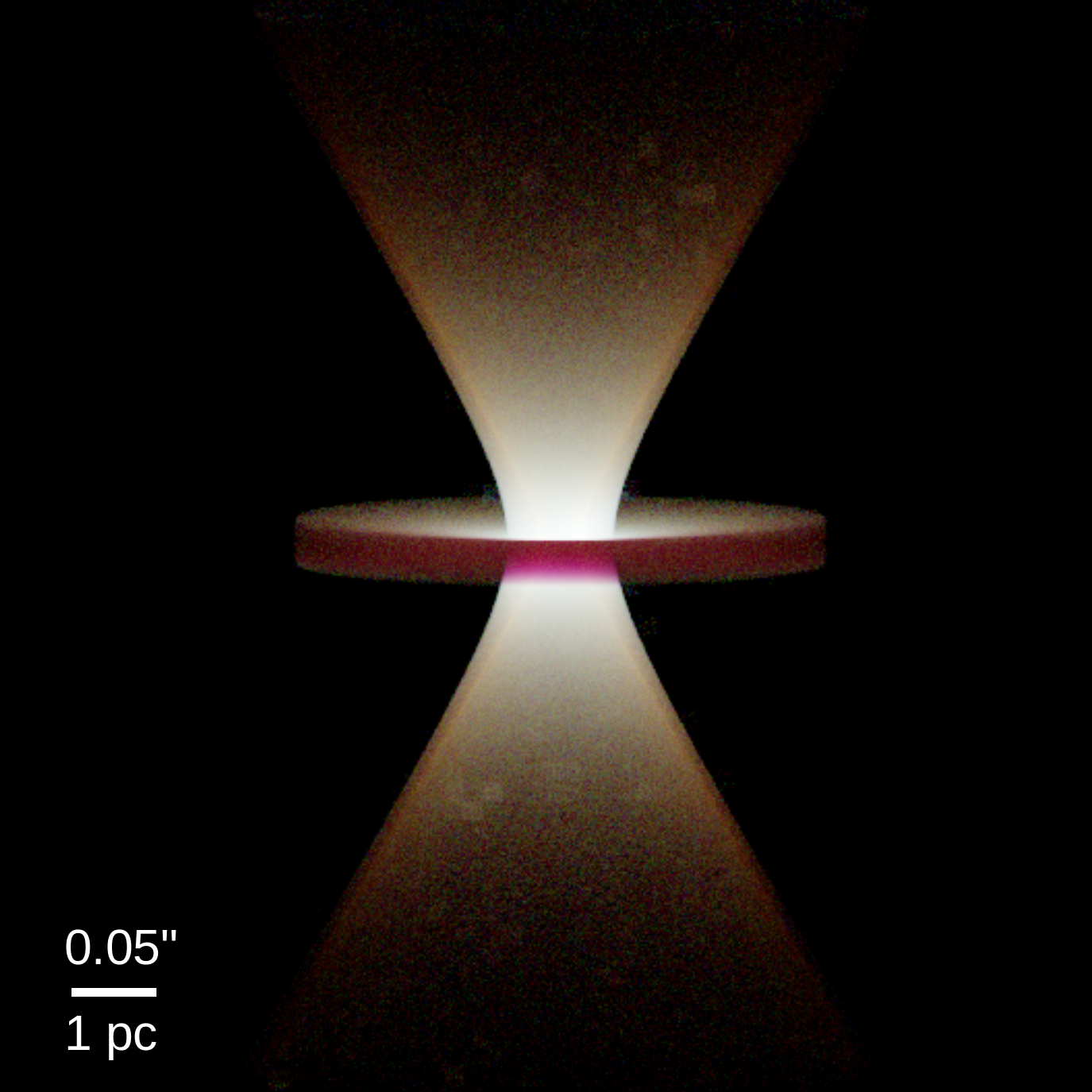}
\caption{Left: schematic of the model geometry, i.e. dust density cut through the \emph{xz} plane: a geometrically thin dusty disc and a polar dusty wind in a form of a hyperboloid shell. The schematic and the values of the parameters correspond to the best model presented in \S \ref{sec:res}. The dust density is constant both in the disc and in the hyperboloid. The angular width of the hyperboloid shell is $1^{\circ}$. Right: a colour composite image (in logarithmic scale) of the best disc$+$hyperboloid model made by mapping the 8, 10 and 13 $\,\micron$ flux images obtained with radiative transfer simulations to the blue, green and red, respectively.}
\label{fig:modrgb}
\end{figure*}

\section{Results and discussion}
\label{sec:res}

\subsection{Model vs. observations: disc$+$polar dust}

\subsubsection{disc$+$hyperboloid}
\label{sec:hyp}

Reaching the convergence when computing the model images is much more computationally expensive than when calculating only SEDs.
Given the long computational time and limited resources, we adopted the following strategy. We computed a sparse grid of model SEDs by varying the parameters of the disc and the hyperboloid. To all the models we applied a foreground extinction screen of $\tau_{9.7}=2$ at the position of the nucleus to account for absorption by dust in the inclined host galaxy disc (see \S ~\ref{sec:dust}). Then we compared these SEDs to the MIDI total flux spectrum and calculated images for those models that provided a satisfactory match. In the next step, we generated synthetic interferometric observations, i.e., we simulated MIDI observations of the model images. This procedure allowed us to narrow down the model parameter space for the next iteration. The procedure was repeated several times until the synthetic interferometric observations of the model reproduced the actual observations well. 

\begin{figure*}
\centering
\includegraphics[trim={0 0.7cm 0 1.9cm},clip,width=1.0\textwidth]{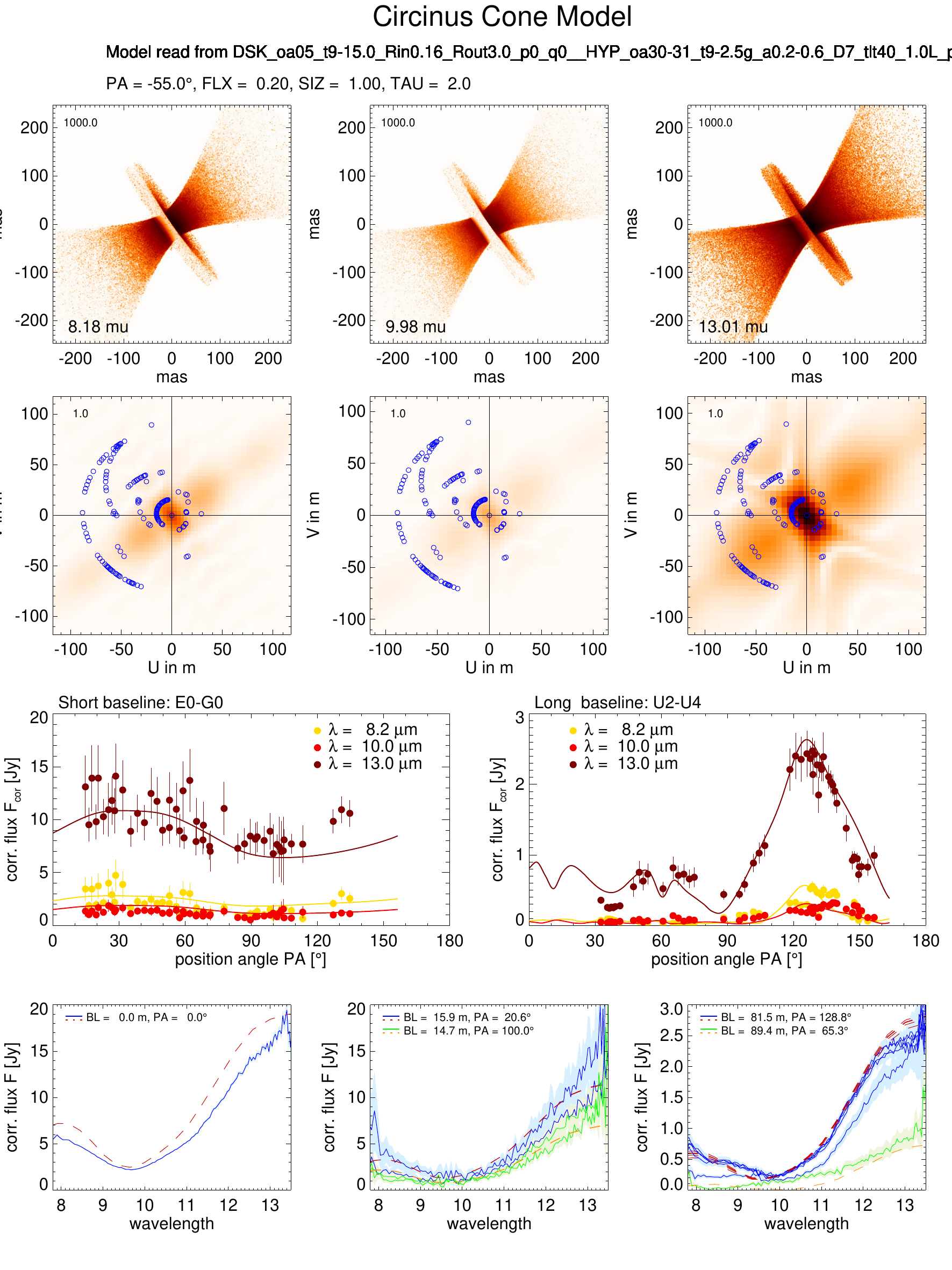}
\caption{The interferometric diagnostic plots for the best disc$+$hyperboloid model. From top to bottom, the rows show: (1) The model images at 8, 10 and 13$\,\micron$. (2) The Fourier transform of the model surface brightness distribution in the \emph{uv} plane with blue circles marking the positions of the interferometric measurements. (3) The correlated fluxes ($F_{\text{cor}}$) of the interferometric measurements (dots) and of the model (lines) as a function of the position angle (PA) of the projected baseline lengths (BL). (4) The total flux spectrum (left) and $F_{\text{cor}}$ spectra extracted from the different baselines and position angles (middle and right); the observed spectra are given in blue and green solid lines, the model spectra in red and yellow dashed lines, with BL and PA indicated within the panels. The signal seen in $F_{\text{cor}}$ vs. PA is well reproduced by the model at both short and long baselines and at all wavelengths.}
\label{fig:hyp}
\end{figure*}

Fig.~\ref{fig:hyp} represents the diagnostic plots we used to compare the models to the data and shows our best disc$+$hyperboloid model with the parameters given further below. This figure contains the model images (first row) and the Fourier transforms of the model surface brightness (second row), but the most important information is in the third row, featuring correlated flux as a function of the position angle, for short and long baselines and at three wavelengths. Here we show correlated fluxes for the two telescope combinations, E0$-$G0 (13 m $<$ BL $<$ 16 m) and UT2$-$UT4 (76 m $<$ BL $<$ 90 m) whose projected baseline lengths remains roughly the same while their position angles cover a wide range due to the rotation of the Earth. This is what makes these two telescope combinations particularly useful: they probe their spatial scales in many different directions. The correlated fluxes of the short E0$-$G0 baselines (third row, left panel) probe larger spatial scales. They show a wide minimum at $\text{PA}\sim90^{\circ}$ indicating extended emission in this direction, which is reproduced by the dusty hyperboloid shell of our model. The correlated fluxes of the long UT2$-$UT4 baselines (third row, right panel), which probe smaller scales, have a prominent peak at $\text{PA}\sim125^{\circ}$. This suggests a very elongated surface brightness of the source, which is poorly resolved along this direction and better resolved along all other directions. In our model, this behaviour is well reproduced by the dusty disc seen almost edge-on. The ``wiggles'' at lower position angles represent non-linear effects due to the interference pattern in the \emph{uv} plane between the disc and the hyperboloid component. In the bottom row of the figure, we see that the model also matches well the correlated flux spectra from the different projected baselines and position angles.

The best model achieved in the described way has the following parameters (also summarised in Table~\ref{tab:param}) of (a) the disc: $\Delta_{\text{dsk}}=5^{\circ}$, $R_{\text{out}}=3\,$pc, $\tau^{dsk}_{9.7}=15$, $M_{\text{dsk}}=102\,M_{\odot}$ and (b) the hyperboloid shell: $\Delta_{\text{hyp}}=30^{\circ}$, $a_{\text{out}}=0.6\,$pc, $\Theta=1^{\circ}$, $\tau^{hyp}_{9.7}=2.5$, $M_{\text{hyp}}=435\,M_{\odot}$ (the inner wall was fixed at $a_{\text{in}}=0.2\,$pc). The total dust mass is dominated by the dust in the polar region: the ratio between the dust masses enclosed in the MIDI aperture is $M_{\text{hyp}}/M_{\text{dsk}}=4.3$, and $M_{\text{pol}}/M_{\text{dsk}}=6.2$ in the VISIR aperture \citepalias[including the polar dust from the VISIR-based model from][]{Stalevski2017}. The dust density is constant ($p=0$) both in the disc and in the hyperboloid. The schematic shown in the left panel of Fig.~\ref{fig:modrgb} corresponds to exactly these values of the parameters, while in the right panel we show a colour composite image of this model. The whole system is inclined towards the observer by $5^\circ$, i.e. the viewing angle measured from the polar axis is $85^\circ$. This is an agreement with \citet{Greenhill2003} who estimated a roughly edge-on orientation of the maser disc. The required accretion disc luminosity is $L_{\text{AGN}}=10^{10} L_{\odot}$, exactly the value inferred by \citet{Tristram2014}. 

The $\tau_{9.7}$ optical depths of the model components can be converted to column densities assuming a relation between the extinction in the $V$-band and column density: $N_{\text{H}}=1.1 \, A_{\text{V}}\times10^{22}\,$cm$^{-2}$ \citep{Maiolino2001, Ricci2014}, where $A_{\text{V}}=1.086 \, \tau_{\text{V}}$ and $\tau_{\text{V}}/\tau_{9.7}$ ratios are given in Table~\ref{tab:dust} for each dust mixture. Based on this, the total edge-on column density of the model is $N_{\text{H}}=2.64\times10^{24}\,$cm$^{-2}$ (including the foreground screen of dust in the host galaxy). 
The line-of-sight column density is somewhat uncertain because the viewing angle is grazing the edge of the dusty disc. Assuming that half of the photons along this direction propagate just above the disc edge and half of the photons just below the edge, we would infer a bit lower value of $N_{\text{H}}=2\times10^{24}\,$cm$^{-2}$. The inferred interval of $N_{\text{H}}$ values of the model is a few times lower than the value found by \citet{Arevalo2014} ($N_{\text{H}}=8\pm2\times10^{24}\,$cm$^{-2}$ ). The difference between the model and observed values can be accounted for by a substantial amount of neutral gas within the dust-free broad line region, as suggested by \citet{Burtscher2016} based on the $N_{\text{H}}/A_{\text{V}}$ ratios.
Note that the column densities inferred observationally from the 9.7$\,\micron$ silicate feature depth and X-rays may not be consistent if the dust distribution is clumpy: $\tau_{9.7}$ would represent absorption integrated in the aperture from which the spectrum was extracted, while X-rays would probe line-of-sight absorption.

The total covering factor of the model, defined by the opening angle of the hyperboloid, is $0.87$. Both disc and hyperboloid have enough optical depth to obscure the accretion disc on their own: the dusty disc would dominate the obscuration for the viewing angles close to edge-on, while for the lower inclinations, the base of the hyperboloid would be responsible for hiding the central source. 
Since polar dust takes a shape of a hollow cone beyond $\sim 1$ pc, the disc$+$wind model allows an unobscured view of the central engine for lower inclinations and thus, it is consistent with the AGN unification-by-inclination scheme, in which obscured and unobscured sources are essentially the same, just seen from a different viewing angle.

\begin{table}
\centering
\caption{A summary of the best model parameters' values.}
\label{tab:param}
\begin{tabular}{ll}
                                                                \hline\hline
\multicolumn{2}{c}{disc parameters}                             \\ \hline
flaring angle                & $\Delta_{\text{dsk}}=5^{\circ}$  \\
outer radius                 & $3$ pc                           \\
edge-on optical depth        & $\tau^{dsk}_{9.7}=15$            \\
dust mass		     & $M_{\text{dsk}}=102\,M_{\odot}$          \\ \hline
\multicolumn{2}{c}{hyperboloid parameters}                      \\ \hline
half opening angle           & $\Delta_{\text{hyp}}=30^{\circ}$ \\
outer wall position          & $a_{\text{out}}=0.6$ pc          \\
edge-on optical depth        & $\tau^{hyp}_{9.7}=2.5$           \\
dust mass		     & $M_{\text{hyp}}=435\,M_{\odot}$         \\ \hline
\end{tabular}

{\it -- Notes:} 
See \S \ref{sec:geo} and Fig.~\ref{fig:modrgb} for the definitions of the parameters.
\end{table}

As mentioned earlier, technical limitations prevented us from computing an extensive grid of models that would densely sample the entire parameter space of the model. For this reason, we are not able to provide quantitative measures of how well the individual parameters are constrained. However, we argue that the dust geometry is rather well constrained, since even relatively small changes of some of the parameters result in a significantly different behaviour of $F_{\text{cor}}$ as a function of PA, clearly inconsistent with the data. For a qualitative estimate of the described effects, in the Appendix we provide several interferometric diagnostic diagrams of the models based on the best one (Table \ref{tab:param} and Fig.~\ref{fig:modrgb}) with variations of some of the parameters. For example, the line of sight needs to be grazing the disc surface or passing by very close to it to reproduce the long-baseline interferometric data (see the effect of inclination change by $\pm5^{\circ}$ in Figs.~\ref{fig:i80} and \ref{fig:i90}). The disc can be a bit more geometrically thick and still consistent with the data, but then lower viewing angles are required, which are inconsistent with an almost-edge-on orientation necessary for the prominent maser disc observed by \citet{Greenhill2003} (Figs.~\ref{fig:oa10i80} and \ref{fig:oa10i90}). Changing the position of the outer wall of the hyperboloid by $\sim30\%$ evidently disrupts the interferometric signature of the dust structure (Figs.~\ref{fig:a0.4} and \ref{fig:a0.8}). Variations of the disc outer radius has a less dramatic, but still visible effect (Figs.~\ref{fig:R1.5} and \ref{fig:R6}). On the other hand, the edge-on optical depths of the disc and the hyperboloid are to a certain degree mutually degenerative, and thus, less well constrained. 
We cannot exclude that the disc could a have higher optical depth and extend further out; this outer part of the disc would thus consist of colder material that would emit at longer wavelengths, in the far-IR (FIR).

In a summary, the overall geometry is well constrained: the data is consistent only with a geometrically thin disc seen almost edge-on and a hyperboloid component perpendicular to the disc.

\subsubsection{disc$+$cone}
\label{sec:con}

\begin{figure*}
\centering
\includegraphics[trim={0 0.7cm 0 1.9cm},clip,width=1.0\textwidth]{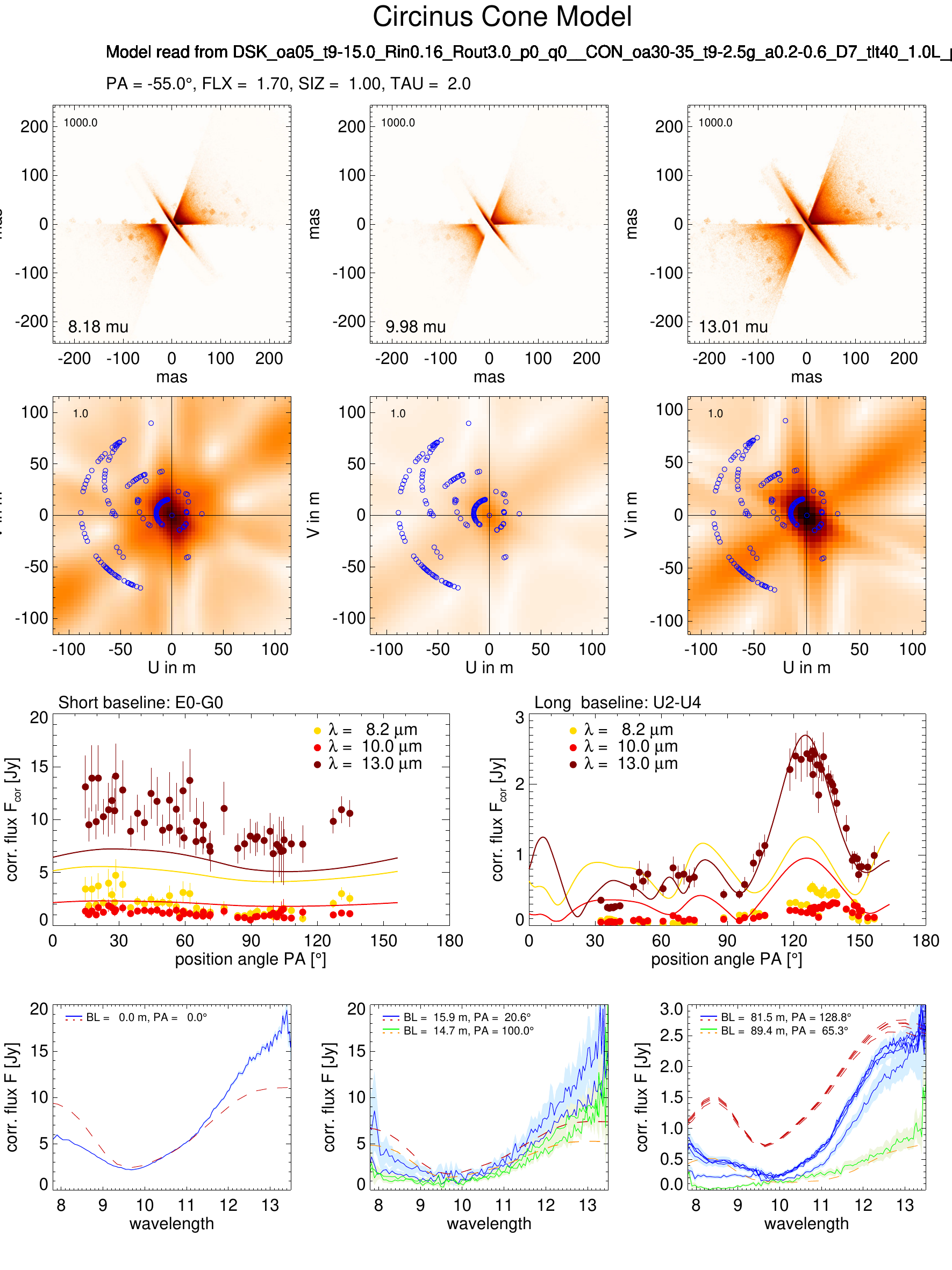}
\caption{The same as in Fig.~\ref{fig:hyp}, but for the disc$+$cone model.}
\label{fig:cone}
\end{figure*}

The interferometric diagnostic plots for the disc$+$cone model with the parameters matching the best disc+hyperboloid are shown in Fig.~\ref{fig:cone}. We see that at the short baseline (third row, left panel), the overall shape of the correlated flux as a function of the position angle is similar to the measured. However, at 13$\,\micron$ the model clearly under-predicts the flux level, while at 8 and 10$\,\micron$ the flux is over-predicted. In other words, the model appears too extended at 13$\,\micron$ and too compact at 8 and 10$\,\micron$. At the long baseline (third row, right panel), the correlated flux is well reproduced by the model at 13$\,\micron$. However, both at 8 and 10$\,\micron$ the model significantly over-predicts the correlated flux at all position angles, i.e. it appears much less extended than it should be. The inadequacy of the model at the short baseline could be potentially somewhat improved by adjusting the amount of the dust and its radial profile in the cone. It is less likely to improve the situation with the 8 and 10$\,\micron$ excess at the long baseline. The reason why the hyperboloid model works well is that its walls are able to hide this excess emission. The cone model fails to do so due to its walls converging towards the centre, and thus, covering only a small portion of the disc. Furthermore, the model spectrum suffers from an excess of $\sim8\,\micron$ emission and a deficit of $\sim13\,\micron$ emission.  While further investigation would be required to fully dismiss the cone geometry, we conclude that the polar dust in the form of hyperboloid shell is clearly favoured by the interferometric data. 

\subsubsection{Polar dust shape vs. theoretical expectations}
\label{sec:modVStheo}

Comparing the hyperboloid and cone models serves well to illustrate how the MIDI data are sensitive even to rather small changes in the geometry of the parsec-scale dust. Additionally, such a well-constrained dust geometry on the smallest scales may provide valuable information for hydrodynamical simulations that aim to produce Circinus-like dust structures \citep[as e.g.][]{Wada2016}. For example, the very geometry, i.e. the fact that the hyperboloid model is preferred over the cone model is indicative of the orientation of the resultant force that is driving the dusty outflow: it must be almost perpendicular to the disc at the base of the wind and then change the orientation farther out to become almost radial, implying that a change of the main wind driver occurs at the given distance. This behaviour is actually expected based on the analytic considerations that take into account gravity of the central black hole, UV radiation pressure of the accretion disc and IR radiation pressure of the dusty disc. For an AGN with the Eddington ratio $\lambda_{\text{edd}}=L_{\text{AGN}}/L_{edd}>0.05$, a dusty disc with a column density $N_{\text{H}}>10^{24}\,$cm$^{-2}$ and sub-Keplerian rotation, the IR radiation pressure would dominate the radiation field at small heights and drive the dust vertically; at larger heights, the radiation pressure of the accretion disc would take over and push the dust away radially (Venanzi et al. in prep). With $N_{\text{H}}>10^{24}\,$cm$^{-2}$ (our model and \citealt{Arevalo2014}), $\lambda_{\text{edd}}\sim0.2$ \citep{Greenhill2003} and a sub-Keplerian maser disc \citep{Greenhill2003}, the AGN in Circinus is expected to be in the described regime of vertical winds at small heights and radial winds at larger heights, corresponding well to the shape of the hyperboloid polar dust in our model.

A very similar geometry of a thin disc and a funnel-shaped shell was proposed by \citet{Elvis2000} to explain the structure of the broad- and narrow-line absorption and emission regions and a number of other spectral features, noting also that with such a model, the overall need for a torus is weaker.

\subsubsection{MIDI-based model vs. VISIR-based model}

The next natural step is to see if the small-scale model that reproduces the MIDI data is consistent with the large-scale (tens of parsecs) model we proposed in \citetalias{Stalevski2017} to explain the VISIR observations. The dusty disc of the MIDI-based model is geometrically thinner, but optically thicker compared to the disc component in the VISIR-based model ($\Delta_{\text{dsk}}$ of $5^{\circ}$ vs. $20^{\circ}$ and $\tau^{dsk}_{9.7}$ of 15 vs 2.5). Another difference is that in the MIDI-based model, there is much more dust in the hyperboloid shell, making it optically thick in MIR. Finally, the MIDI data require that the half-opening angle of the hyperboloid is $10^{\circ}$ smaller than in the large-scale model. However, the spatial scales that are probed by MIDI correspond to the unresolved core in the VISIR images. In \citetalias{Stalevski2017}, our goal was to reproduce the large-scale MIR morphology, so we explored only a limited range of the parameters of the small-scale region which is unresolved in VISIR images.

In Fig.~\ref{fig:SED}, we compare our VISIR-based model (solid black line) from \citetalias{Stalevski2017} and the MIDI-based model from this work (dash-dotted red line) with the observed SED taken from \citetalias{Stalevski2017}. For all the observations, we measured the fluxes consistently in a $4\arcsec$ diameter aperture (corresponding to the total VISIR aperture) and a $0.4-0.6\arcsec$ diameter aperture (corresponding to the unresolved core in VISIR and the MIDI total flux spectrum aperture). More details on the assembled SED can be found in Section~2 and Table 1 of \citetalias{Stalevski2017}. In Fig.~\ref{fig:SED}, we can see that the MIDI-based model offers a good match to the MIDI spectrum and to the MIR photometry from the corresponding apertures. In particular, the MIDI model is matching very well the small aperture MIR photometry, consistent with the unresolved core emission in VISIR images. However, it is evident that the model under-predicts the hot dust emission in the near-IR (NIR). This is due to the hyperboloid shell which is optically thick in the IR, so it efficiently self-absorbs its own NIR emission, but also blocks the view of the innermost hot region of the dusty disc.

\subsubsection{Tensions and possible solutions}

We identified the following potential tensions between the MIDI- and VISIR-based models: (a) the half-opening angle of the hyperboloid is $10^{\circ}$ smaller than in the VISIR-based model, (b) the MIDI-based model under-predicts the NIR emission due to the self-absorption and obscuration and (c) the hyperboloid shell is optically thick in the radial direction, which would prevent the light from the central source to sufficiently heat up the dust far away from the nucleus. We offer the following solution for these tensions.

For the opening angle of the VISIR-based model, we adopted the outer half-opening angle obtained from kinematic modeling of the narrow line region by \citet{Fischer2013}. However, they also provide an inner half-opening angle of the cone of $36^{\circ}$. Assuming this value, the discrepancy would be only $6^{\circ}$. Furthermore, the shape of the polar dust could simply be such that the hyperboloid opening angle is narrower close to the base of the outflow and wider at larger scales to match the opening angle of the ionisation cone. It is beyond the scope of this work to investigate the reasons for this; we speculate this could happen due to the change of the ambient pressure and/or the change of the dominant radiation that is driving the dusty wind, as described in \S \ref{sec:modVStheo} \citep[see also][]{Chan-Krolik2016, Chan-Krolik2017}. Additionally, the actual dust geometry may be more complex, with varying optical depth along the different directions and with some dust within the cone, resulting in requiring narrow opening angle in our simple model.

\begin{figure*}
\centering
\includegraphics[width=1.0\textwidth]{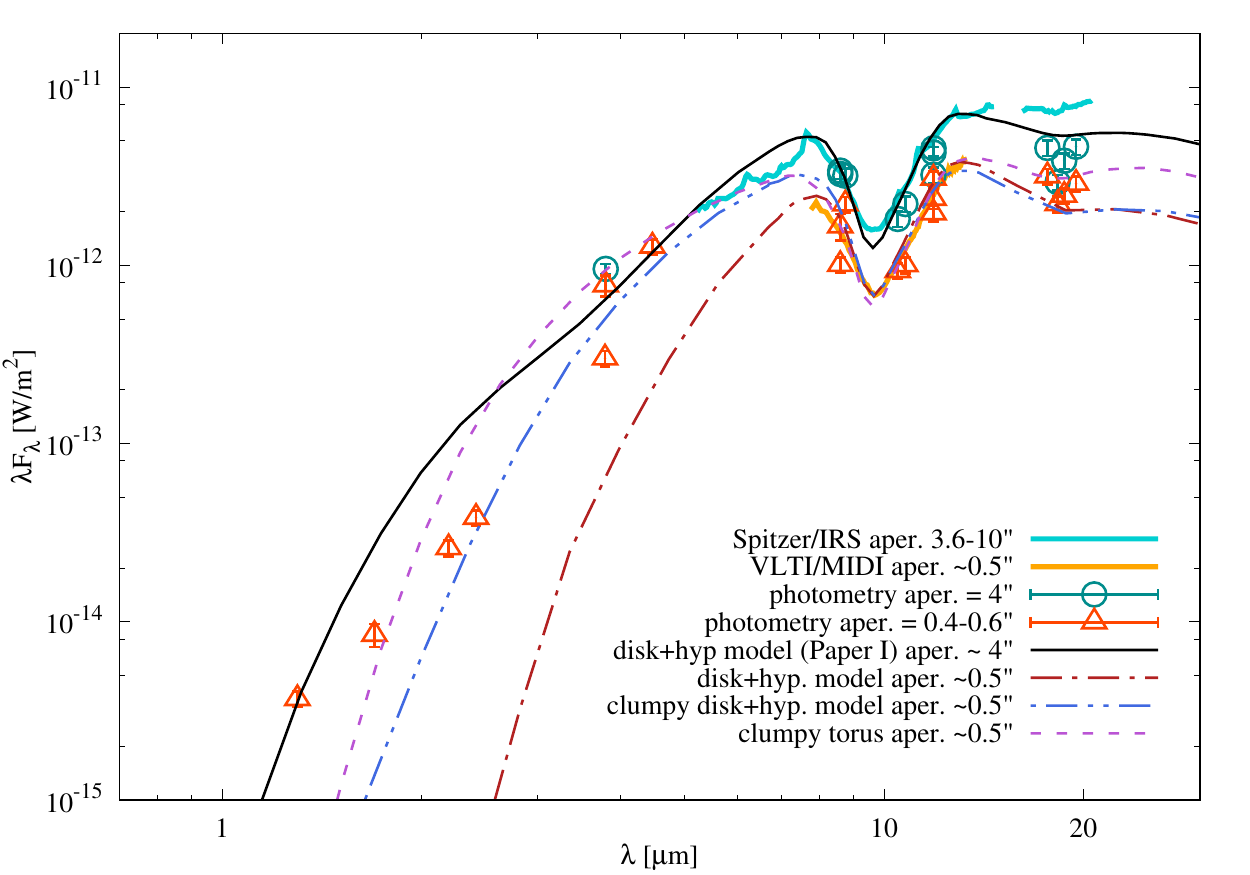}
\caption{Comparison of the observed SED with the VISIR-based model SED from \citetalias{Stalevski2017} (solid black line), the MIDI-based models from this work (a smooth model in dash-dotted red line and a clumpy model in dash-double-dotted blue line), and a representative clumpy torus model (dashed orchid line). Large-scale aperture photometry ($4\arcsec$) is shown in green down-pointing triangles, while photometry extracted from apertures comparable or smaller than the resolution limit of VISIR ($0.4-0.6\arcsec$) is marked by red up-pointing triangles. The aperture of the Spitzer/IRS spectrum ($\geq3.6\arcsec$) is comparable to the total aperture of VISIR in $5.2-14.5\,\micron$ range, while significantly larger at longer wavelengths. The MIDI spectrum was extracted using a $\sim0.54\arcsec\times0.52\arcsec$ aperture and hence corresponds to the unresolved nucleus with VISIR.}
\label{fig:SED}
\end{figure*}

The two remaining tensions, i.e., the lack of the NIR emission and optical thickness of the hyperboloid shell, could be naturally overcome if the dust in the hyperboloid shell is clumpy. The clumpy models require much more computational memory in order to properly resolve the individual dust clouds, so in our modeling we started with a smooth dust distribution, which proved to work well. However, a radiation-driven dusty wind is likely to result in an turbulent, filamentary or clumpy medium \citep[e.g.][]{Wada2012}. Such an inhomogeneous outflow could allow a certain amount of hot NIR emission to come through to the observer, as well as to provide enough dust-free lines along which photons could escape in the radial direction of the hyperboloid shell and illuminate the dust farther out to produce the large-scale MIR morphology. The computational resources at our disposal prevented us from doing an in-depth study of the clumpy models; however, even with a limited amount, we were able to show that the proposed solution is plausible. We calculated a small amount of models with exactly the same geometry, but with the hyperboloid shell composed of a number of spherical clumps, with a several combinations of the number of clumps and their size that result in the volume filling factors of the shell of 0.1, 0.15, 0.2, 0.25. We found that the models with the filling factor of 0.25 offer a good match to the observations. The SED of this model is shown in Fig.~\ref{fig:SED} as dash-double-dotted blue line. We see that this clumpy model SED is consistent with the smooth model in the MIR range, but has a higher lever of hot dust emission, consistent with the most of the NIR data, under-predicting significantly only the emission shortward of $\sim2\,\micron$. 

\begin{figure*}
\centering
\includegraphics[trim={0 0.6cm 0 2.0cm},clip,width=1.0\textwidth]{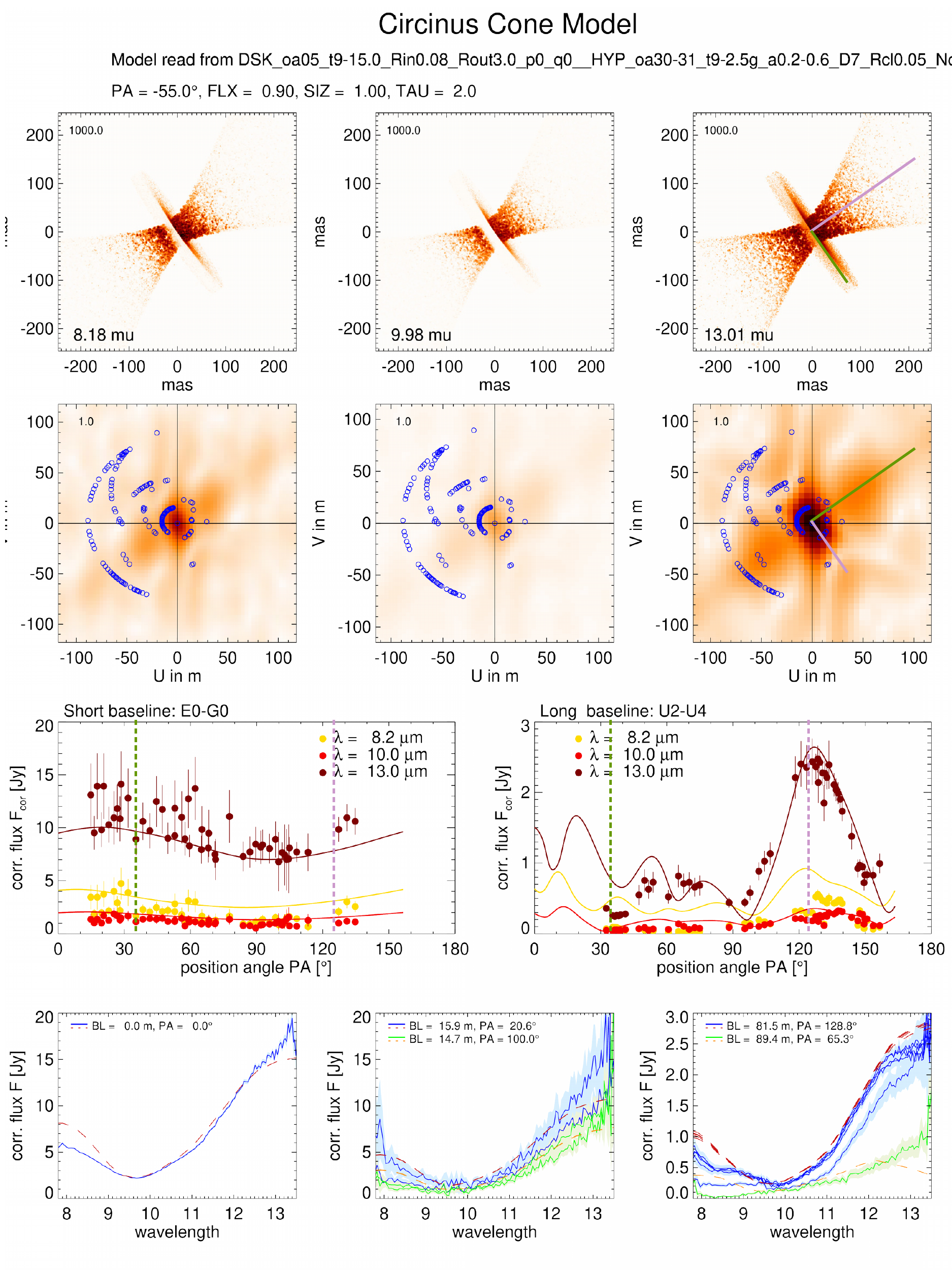}
\caption{The same as in Fig.~\ref{fig:hyp}, but for the disc$+$clumpy hyperboloid model. Green and light magenta lines are marking the position angles of the projected baselines which probe the dust structure in the two perpendicular directions, along the disc and in the polar direction; note that these two directions are flipped in the \emph{uv} plane due to the Fourier transform.}
\label{fig:clumpy}
\end{figure*}

In Fig.~\ref{fig:clumpy}, we show the interferometric diagnostic plots for this clumpy model. We see that the clumpy model also reasonably reproduces the correlated flux dependency on the position angle for the short baseline (third raw, left panel). For the long baseline (third row, right panel), the clumpy model also fits well the main feature of the correlated flux: the peak at the position angle $\text{PA}\sim125^{\circ}$ at 13$\,\micron$. Only at 8$\,\micron$ the model deviates significantly, predicting higher correlated fluxes than the measured at most of the position angles. In other words, the model appears less extended at this wavelength than it should. This mismatch at 8$\,\micron$ could likely be reduced by further adjustment of the clumpy model, including a number of different random realisations of the clumps' positions. Here we simply aimed to demonstrate that clumpiness can help alleviate the mentioned issues, and did not purse the investigation of the clumpy model further due to the numerically expensive calculations.

Additionally, the bolometric accretion disc luminosity required for the VISIR-based model is $\sim30\%$ higher than the one in the MIDI-based model. This could be within the uncertainties of the model, which are difficult to quantify since we were not able to explore the full parameter space. Nevertheless, if the discrepancy is not simply due to the model uncertainties, this would imply the existence of an additional heating source on the larger scales. A young stellar population within 16 pc of the nucleus is found to have $L^{*}=2.3\times10^{8} L_{\odot}$ \citep{MullerSanchez2006}, corresponding to only $2.3\%$ of the AGN luminosity inferred by our model. However, as there is evidence of significant extinction toward the nucleus, the intrinsic stellar luminosity could be much higher; an upper limit is difficult to determine \citep{MullerSanchez2006}. Furthermore, \citet{Maiolino1998} found that the luminosity of young stars within 200 pc of the nucleus is of order $10\times10^{10} L_{\odot}$, comparable to the AGN luminosity. Another mechanism that could account for the missing luminosity is collisional dust heating by electrons or protons, assuming dust is mixed with a hot gas or entrained in an ionised wind \citep{Bocchio2013}.

As final step in this test, ideally we would calculate the MIR images of the clumpy MIDI-based model extending out $40\,$pc and produce the synthetic VISIR observations to see if the same large-scale morphology is reproduced. However, such a model would require a large number of very small clumps. While we could ensure that the clumps are well resolved in the small-scale model, the large-scale model would require many more clumps, making the simulations prohibitively expensive in the terms of required computational memory. Thus, while the technical limitations prevented us from preforming a conclusive test, we have shown that the disc$+$hyperboloid model, that is reproducing well the interferometric observations, is also reasonably consistent with the model we presented in \citetalias{Stalevski2017} that is able to explain the MIR morphology seen in the VISIR images. The potential tensions likely arise due to the more complex structure on the small scales and can be mitigated under reasonable assumptions, as we have argued.

\subsection{Model vs. observations: dusty torus}

\begin{figure*}
\centering
\includegraphics[trim={0 0.7cm 0 1.9cm},clip,width=1.0\textwidth]{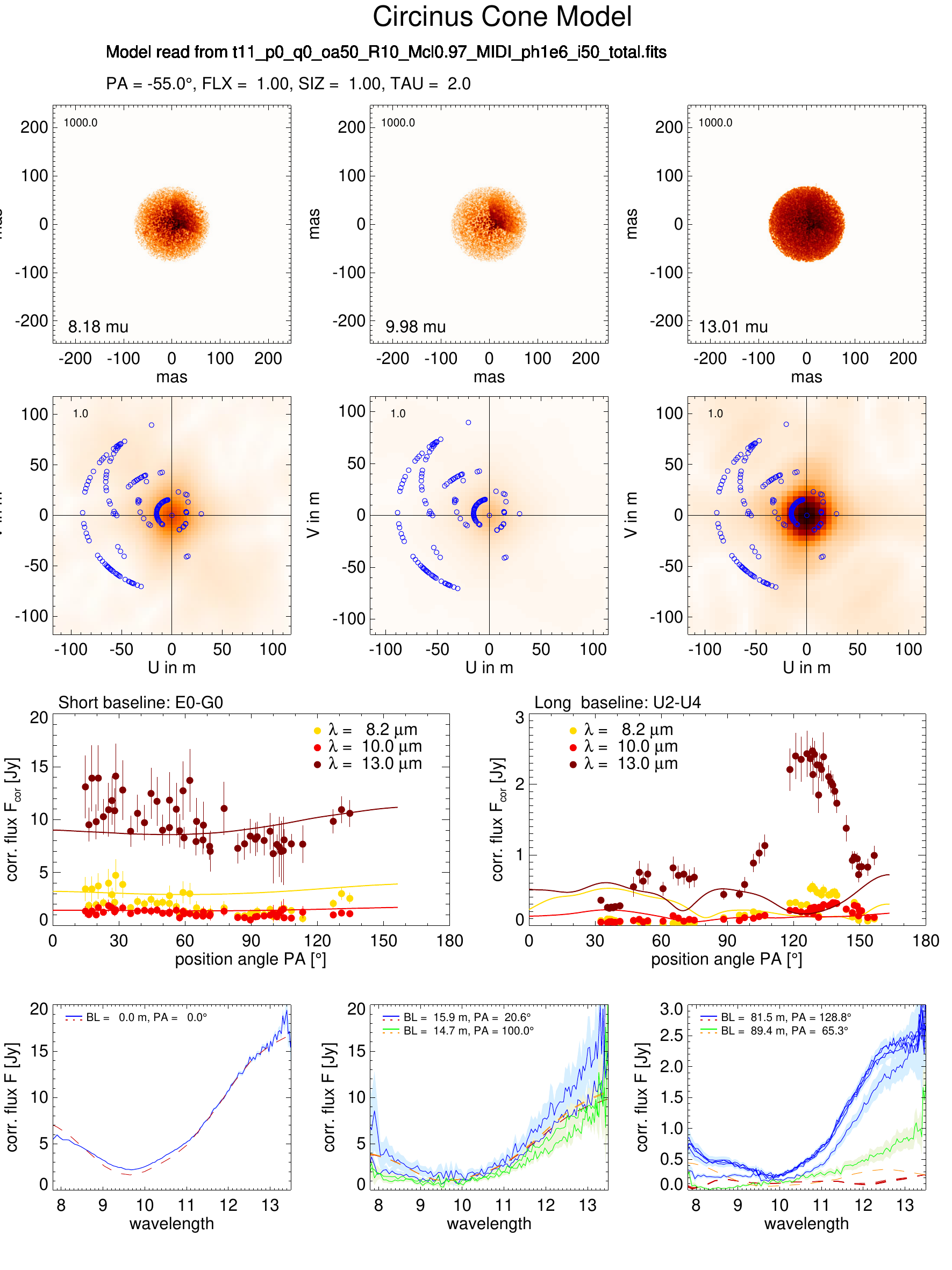}
\caption{The same as in Fig.~\ref{fig:hyp}, but for the representative torus model. While the torus model can provide a good fit to the total flux spectrum, it fails to reproduce the interferometric data. The parameters of the shown model are: $\tau^{dsk}_{9.7}=11$, $\Delta_{\text{tor}}=50^{\circ}$, $R_{\text{out}}/R_{\text{in}}=10$, $p=0$, $q=0$, $i=50^{\circ}$.
}
\label{fig:torus}
\end{figure*}

We used the whole \textsc{SKIRTOR} library of 19200 SEDs (attenuated by a foreground extinction screen of $\tau^{dsk}_{9.7}=2$ to account for absorption by dust in the host galaxy) to fit the MIDI spectrum and then computed images only for those models that were a good match. Among those models, high optical depth is preferred ($\tau^{dsk}_{9.7}=11$), constant radial distribution of the clouds ($p=0$) and half-opening angles in the range of $\Delta_{\text{tor}}=50-80^{\circ}$. The distribution of the clouds with the polar angle ($q$ parameter in the density law) and the outer radius of the torus are not well constrained. While a number of these models provide a good fit to the SED (an example shown with dashed orchid line in Fig.~\ref{fig:SED}), none of them are consistent with the interferometric data. In Fig.~\ref{fig:torus} we show the interferometric diagnostic plots for a representative torus model, which apart from the good fit to the spectrum, has the opening angle that is matching the opening angle of the ionisation cone. In the short baseline (third row, left panel), we see that the correlated flux remains flat at the smaller position angles and only marginally increases at larger position angles. This means that the model appears almost symmetric, without significant elongation in the polar direction. The situation is much worse at the long baselines: the torus model completely fails to reproduce any feature of the correlated flux as a function of the position angle. This is because the model lacks a component that could resemble a very elongated surface brightness distribution that is needed to match the peak of the correlated flux at the position angle of $\text{PA}\sim125^{\circ}$. 

We tested if the model would perform better if the torus is combined with an additional disc component with exactly the same parameters as the one from the best disc$+$hyperboloid model. The results remain essentially unchanged, since the dust clouds from the higher altitudes effectively completely hide the disc from our view. 
A torus combined with a hyperboloid completely fails in the long baseline for the same reason stated above; at the short baseline, the torus dust clouds interfere with the base of the hyperboloid outflow and a roughly sinusoidal signature in the short baseline is still present, but smoothed-out. In other words, parsec-scale cannot host a thick disc, torus or another similar, geometrically thick structure. The interferometric signatures of the other torus models with a good fit to the spectrum do not differ significantly from the one shown in Fig.~\ref{fig:torus}.

If observational constrains from other wavelengths are taken into account, a case against a geometrically thick disc/torus is even stronger. Namely, if we consider only the torus models with opening angles that would match the opening angle of the ionisation cone, a good fit to the MIDI spectrum is achieved exclusively by lower viewing angles ($i\leq60^{\circ}$). Such low inclinations are not consistent with the observed maser disc, which requires a close to edge-on viewing angle \citep{Greenhill2003}.

We conclude that, while the torus model can successfully match the SED, it is firmly ruled out as it fails to reproduce the interferometric data. Therefore, the existence of a parsec-scale, geometrically thick, warm dust structure is not compatible with the observational data, and the interpretation that the polar component seen in MIDI is actually the inner illuminated part of the torus is ruled out. The absence of a dusty torus in Circinus was also proposed by \citet{Mezcua2016} who argued that the required obscuration of the central source and collimation of its emission in the ionisation cone could be caused by dust lanes of the host galaxy on $\sim10$ pc scales. We note that we do not exclude the existence of additional material beyond the outer radius of the warm dusty disc, but it could be only cold material, emitting primarily in the FIR, and geometrically thin (or consisting of a relatively small number of clumps), so that it does not contribute significantly, nor absorb too much of the MIR emission.

\subsection{Comparison with other works}

\begin{figure*}
\centering
\includegraphics[width=0.49\textwidth]{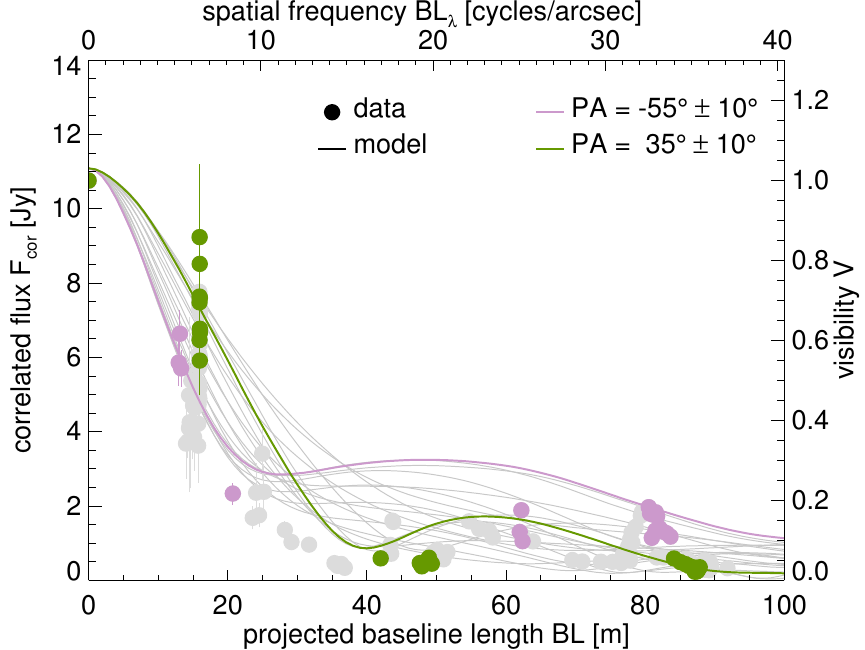}
\includegraphics[width=0.5\textwidth]{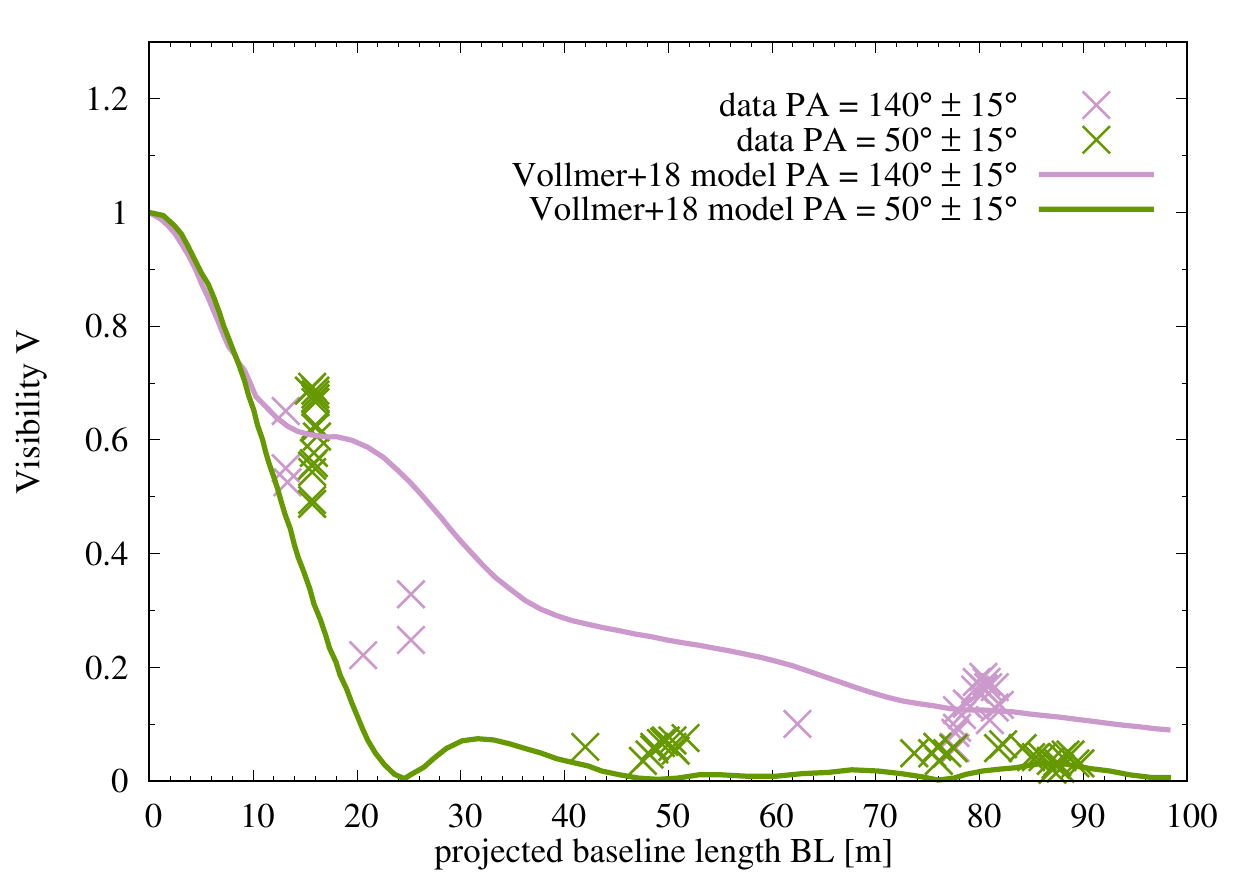}
\caption{Left: Correlated fluxes ($F_{\text{corr}}$, left axis) or visibilities ($V = F_{\text{corr}}/\overline{F}_{\text{tot}}$, right axis) at 12$\,\micron$ of the data (points) and the model (lines) as a function of the projected baseline length (BL, bottom axis) or spatial frequency (top axis). Highlighted are the two position angles (PA) marked in Fig.~\ref{fig:clumpy} with green ($35^{\circ}$) and light magenta ($-55^{\circ}\equiv125^{\circ}$) lines. The uncertainties of the measurements for BL $>$ 30 m are smaller than the plot symbols. The apparent scatter of the data at some baseline lengths is mostly due to measurements at different position angles. Right: the equivalent plot from \citetalias[Fig.~20]{Vollmer2018}. Note a different convention in measuring PAs: $-55^{\circ}$ in our plot would correspond to $125^{\circ}$ in \citetalias{Vollmer2018}. Due to the different configuration of our model, we choose to probe it in slightly different, but still consistent range of PAs.}
\label{fig:PAvsBL}
\end{figure*}

To the date, there are only two works in the literature that focus on modeling of the interferometric MIDI data of Circinus: \citet{Tristram2014}, on which our work is based on, and \citet[hereafter V18]{Vollmer2018}. In \citetalias{Vollmer2018}, the authors considered turbulent gas discs and demonstrated that magneto-centrifugal force, with an assistance of radiation pressure, is able to launch a wind in the polar direction. This wind can carry away the angular momentum and allow the disc to become geometrically thin. The authors showed that such a model is in general consistent with a number of AGN observables and applied their model to the Circinus by assuming a dust structure consisting of a thick disc, a thin disc, a cone-like wind and a puffed-up region on the inner, thin disc. The transition between the thick and the thin disc occurs at $0.53\,$pc, which is also the wind-launching radius. A puff-up, located close to the sublimation radius, is introduced to provide the necessary additional NIR emission. Our model, which is purely phenomenological, assumes only a single disc and a wind launched close to the sublimation radius. \citetalias{Vollmer2018} assume parabolic shape for the polar component. In \citetalias{Stalevski2017}, we rejected the parabolic shape since we demonstrated that such a geometry would produce a warped bar with changing orientation at large scales, inconsistent with a straight bar seen in the VISIR images. Instead, the polar component in our model is in the shape of hyperboloid. Thus, while globally similar, no direct comparison between the parameters of the two models is meaningful. However, we can still consider how well the two models reproduce the MIDI data. For this purpose we can compare visibilities as a function of the projected baseline shown in the left panel of Fig.~\ref{fig:PAvsBL} with the equivalent plot in Fig.~20 from \citetalias{Vollmer2018} here reproduced in the right panel. 

In Fig.~\ref{fig:PAvsBL}, we over-plot in grey the data and model values for all the PAs, but for clarity we highlight in colour only values for the two position angles: $35^{\circ}$ (green) and $-55^{\circ}$ ($\equiv125^{\circ}$) (light magenta). These two PAs are particularly useful since they probe the structure in the two perpendicular directions, along the disc and in the polar direction (marked with the same colours in Fig.~\ref{fig:clumpy}). We see that our model reproduces well the observed change of $F_{\text{corr}}$ as a function of BL: at short baselines, the polar wind component (probed by $\text{PA}=125^{\circ}$) is better resolved than the disc component (probed by $\text{PA}=35^{\circ}$), while at longer baselines it is the opposite. 

The model of \citetalias{Vollmer2018} is in a good agreement with the observed visibilities at long baselines. However, at short baselines (BL $<$ 25 m), their model does not match the data so well and the visibilities of the polar component (probed by $\text{PA}=140^{\circ}$ in their case) are above or overlap with the visibilities of the disc component (probed by $\text{PA}=50^{\circ}$ in their case). In other words, the disc-like component is well reproduced by the \citetalias{Vollmer2018} model, but the polar component does not appear to be sufficently prominent.

\section{Summary and conclusions}
\label{sec:sum}

The MIR emission of number of nearby AGN has been resolved recently. Surprisingly, these observations revealed that a major fraction of MIR emission is coming from the region extended in the direction of the AGN polar axis, in contrast to what is expected from the canonical dusty torus oriented in the equatorial plane. This motivated works that put forth an idea that the polar MIR emission is associated with dusty winds driven by radiation pressure, leading towards a new paradigm for the structure of dust in AGN.
Alternatively, the polar elongation might be produced by a standard dusty torus seen from a favourable viewing angle and/or surrounded by ambient, host galaxy dust. To make further progress and test the different scenarios, case studies of objects with high intrinsic resolution are of crucial importance.
The AGN in the Circinus galaxy represents one of the best examples with detected polar dust emission, from parsec-scale to tens of parsecs scale. 

We performed a detailed modeling of the Circinus AGN, constructing a model based on observations across a wide range of wavelengths and spatial scales. In \citetalias{Stalevski2017}, we proposed a model consisting of a compact dusty disc and a large-scale hollow dusty cone illuminated by a tilted accretion disc. We showed that this model is able to reproduce well the observed $40\,$pc scale MIR morphology of Circinus and its entire IR SED. In the work presented here, we refined the model further by following insights from the interferometric VLTI/MIDI observations, which probe the parsec-scale region. Employing a state-of-the-art 3D radiative transfer code based on the Monte Carlo technique, we produced the images of our model consisting of a dusty disc seen almost edge-on and a dusty polar outflow in the form of a hyperboloid shell. Using these images we simulated interferometric observations of the model to compare it to the MIDI data. We tested in the same way if the canonical dusty torus with a favourable configuration can reproduce the data. From our investigation, we conclude the following:

\begin{itemize}

 \item The disc$+$hyperboloid model is able to reproduce well the interferometric observations, i.e. the shape of the correlated flux as a function of the position angle, at all wavelengths and all baselines. Such a shape of a dusty wind in the polar region is matching the theoretical expectations based on the inferred values of the column density, the Eddington ratio and sub-Keplerian rotation for AGN in Circinus.
 
 \item The model is reasonably consistent with the large-scale model from \citetalias{Stalevski2017}. Some minor adjustments are needed, such as clumpiness of the hyperboloid shell to account for the missing NIR emission.

 \item The clumpy dusty torus model can provide a good fit to the integrated IR SED, but it fails to reproduce the resolved interferometric data. The interpretation that the polar component seen in the VLTI/MIDI data is actually the inner illuminated part of the torus surface is ruled out.
  
 \end{itemize}

\emph{We conclude that the existence of a parsec-scale, geometrically thick, warm dust structure is not compatible with the observations. Instead, the data is consistent only with a dusty structure consisting of a thin disc seen almost edge-on and a wind in the polar direction in the shape of a hollow hyperboloid.}
 
Our results reinforce calls for caution when using the dusty tori models to interpret the IR data of AGN. Further study of polar dust emission in larger samples is necessary to constrain its properties and assess its ubiquity in the whole AGN population. To meet this goal, new models of AGN dust emission are needed. We propose the here presented disc$+$hyperboloid wind model of the AGN in the Circinus galaxy to serve as a prototype for the dust structure in the AGN population with polar dust.

\section*{Acknowledgements}

MS acknowledges support by the Ministry of Education, Science and Technological Development of the Republic of Serbia through the projects Astrophysical Spectroscopy of Extragalactic Objects (176001) and Gravitation and the Large Scale Structure of the Universe (176003). An early stage of this research was supported by FONDECYT through grant No.\ 3140518 and by the Center of Excellence in Astrophysics and Associated Technologies. Part of this work has been performed under the Project HPC-EUROPA3 (INFRAIA-2016-1-730897), with the support of the EC Research Innovation Action under the H2020 Programme; in particular, MS gratefully acknowledges the support and hospitality of Vassilis Charmandaris from the Department of Physics, University of Heraklion, Crete and the computer resources and technical support provided by GRNET. MS is also grateful for the hospitality and support by ESO headquarters in Santiago, Chile, where part of this work was conducted, funded through the Scientific Visitor program. This work made use of ESO Chapman computer; we thank its technical personnel for support and other users for their patience. Powered@NLHPC: This research was partially supported by the supercomputing infrastructure of the NLHPC (ECM-02). This research made use of Astropy, a community-developed core Python package for Astronomy \citep{Astropy2013}. DA acknowledges support from the European Union’s Horizon 2020 ERC Starting Grant DUST-IN-THE-WIND (ERC-2015-StG-677117) and the Innovation programme under the Marie Sklodowska-Curie grant agreement no. 793499 (DUSTDEVILS). We thank the anonymous referee for careful reading and helpful comments which improved the clarity of the paper.


\bibliographystyle{mnras}
\input{circinus-II_v5_16012019-1_accepted_arxiv.bbl}

\appendix

\section{Impact of parameter variation on the interferometric observables}

\begin{figure*}
\centering
\includegraphics[trim={0 0.7cm 0 1.9cm},clip,width=1.0\textwidth]{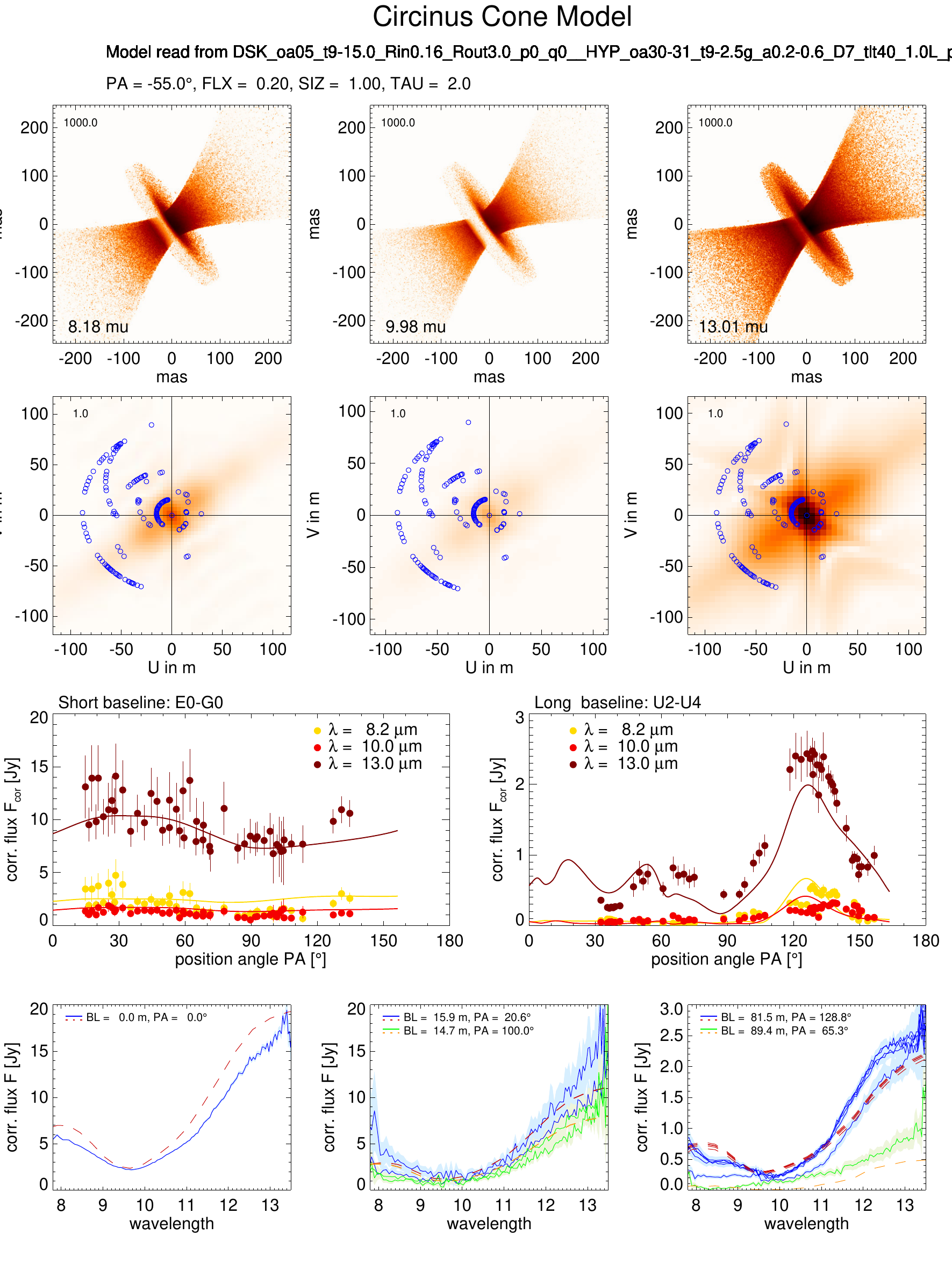}
\caption{The same as in Fig.~\ref{fig:hyp}, but for the viewing angle $i=80^{\circ}$.}
\label{fig:i80}
\end{figure*}
\begin{figure*}
\centering
\includegraphics[trim={0 0.7cm 0 1.9cm},clip,width=1.0\textwidth]{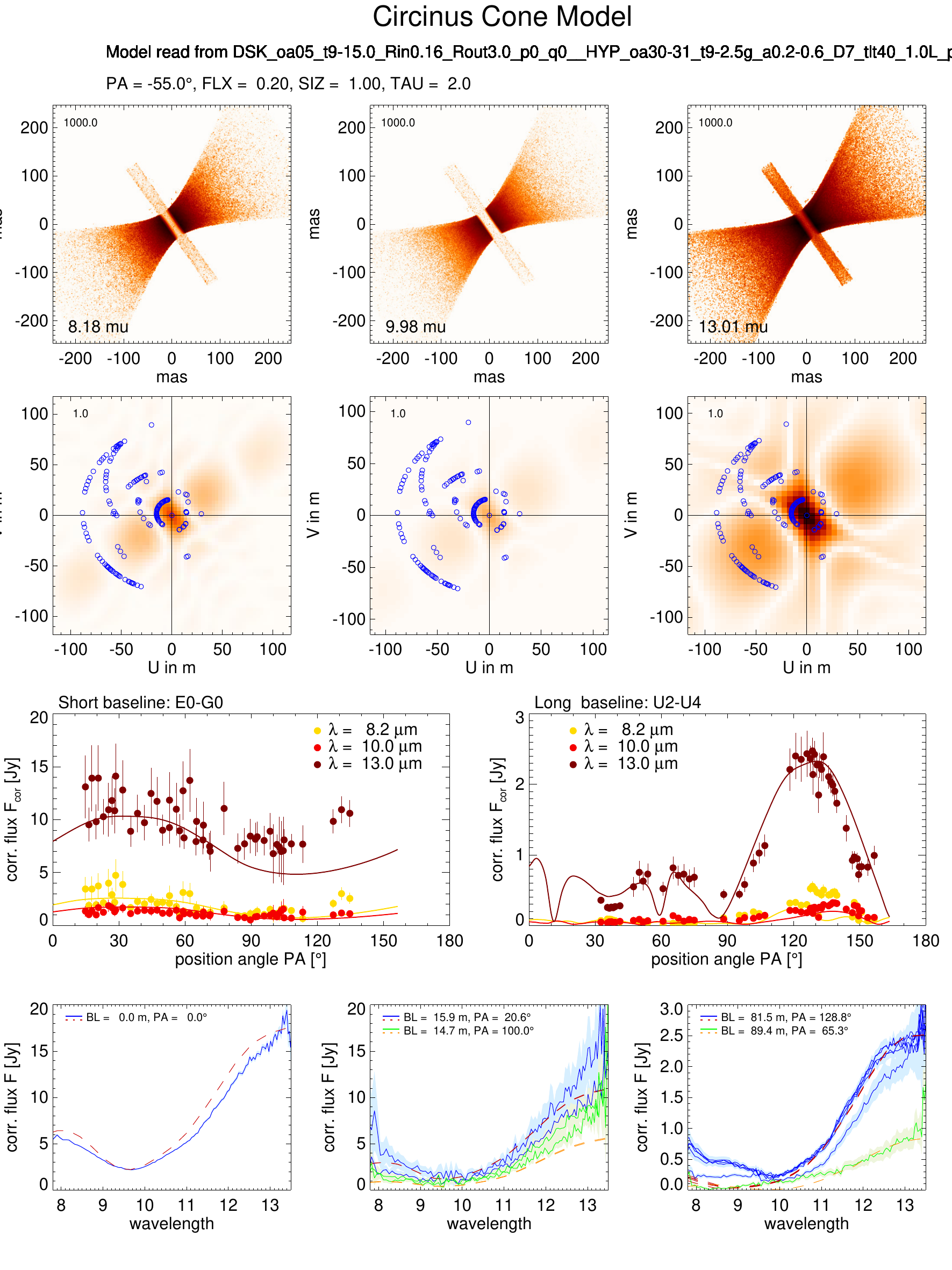}
\caption{The same as in Fig.~\ref{fig:hyp}, but for the viewing angle $i=90^{\circ}$.}
\label{fig:i90}
\end{figure*}
\begin{figure*}
\centering
\includegraphics[trim={0 0.7cm 0 1.9cm},clip,width=1.0\textwidth]{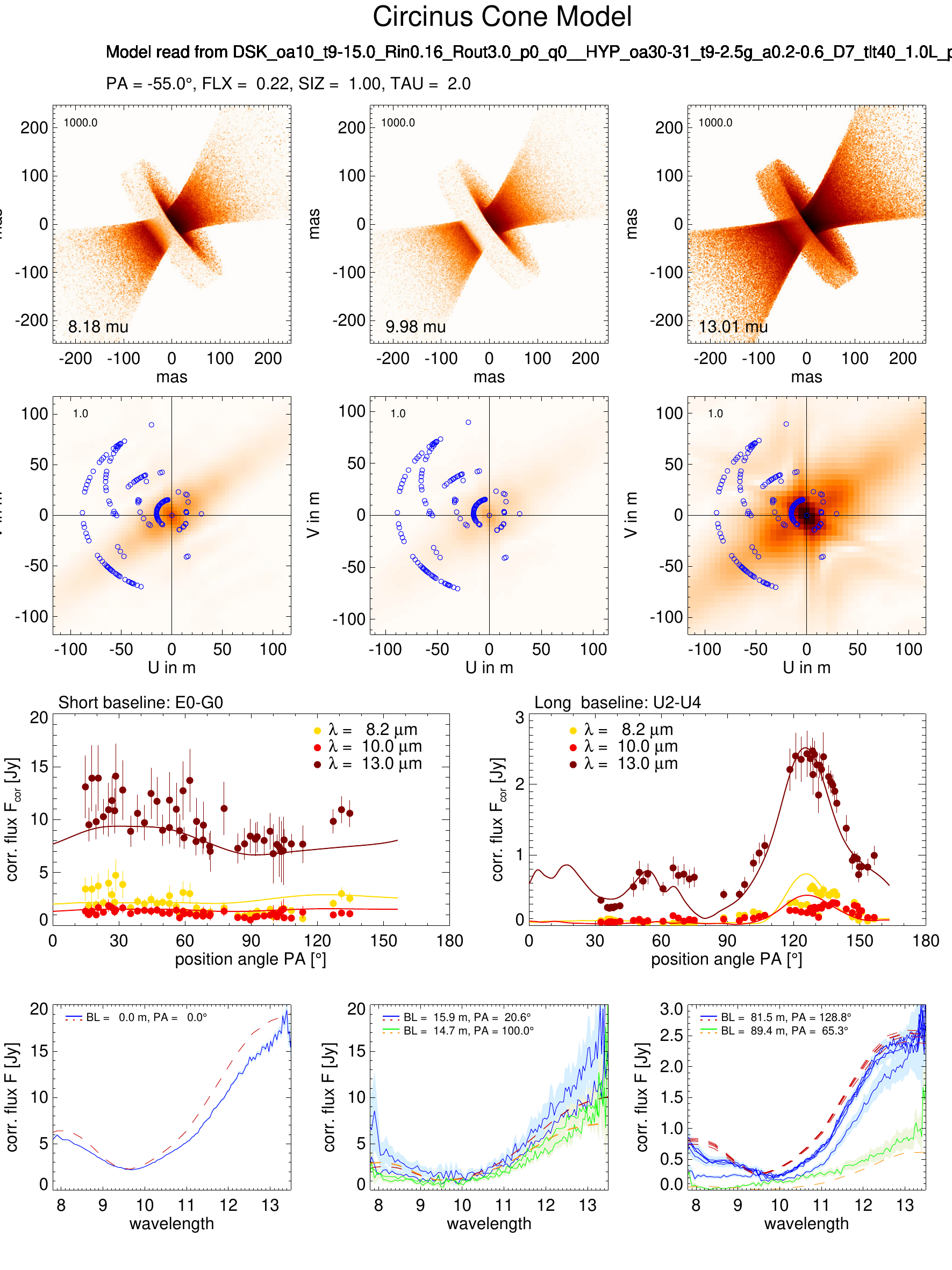}
\caption{The same as in Fig.~\ref{fig:hyp}, but for the disc flaring angle $\Delta_{\text{dsk}}=10^{\circ}$ and the viewing angle $i=80^{\circ}$.}
\label{fig:oa10i80}
\end{figure*}
\begin{figure*}
\centering
\includegraphics[trim={0 0.7cm 0 1.9cm},clip,width=1.0\textwidth]{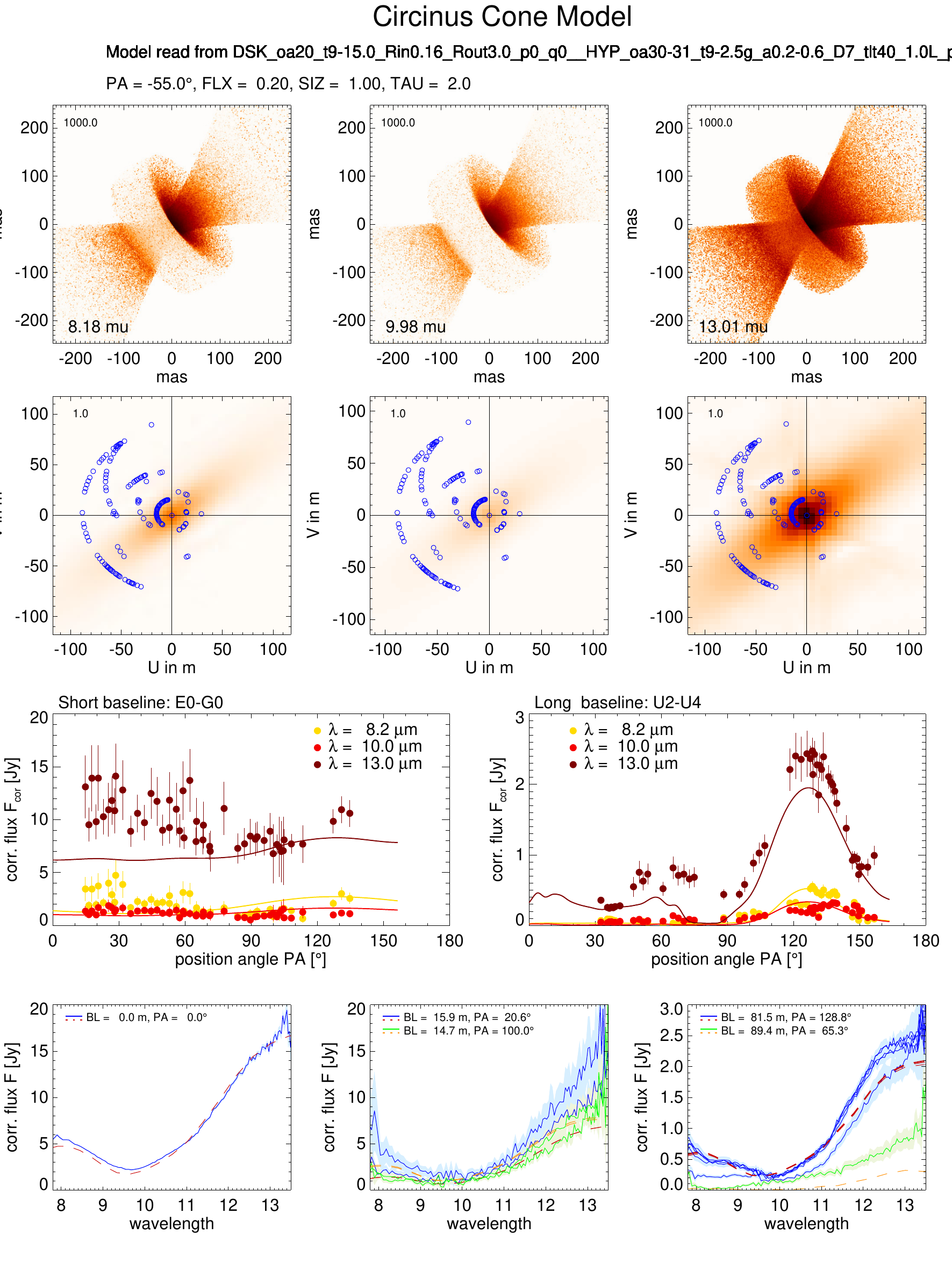}
\caption{The same as in Fig.~\ref{fig:hyp}, but for the disc flaring angle $\Delta_{\text{dsk}}=20^{\circ}$ the viewing angle $i=70^{\circ}$.}
\label{fig:oa10i90}
\end{figure*}
\begin{figure*}
\centering
\includegraphics[trim={0 0.7cm 0 1.9cm},clip,width=1.0\textwidth]{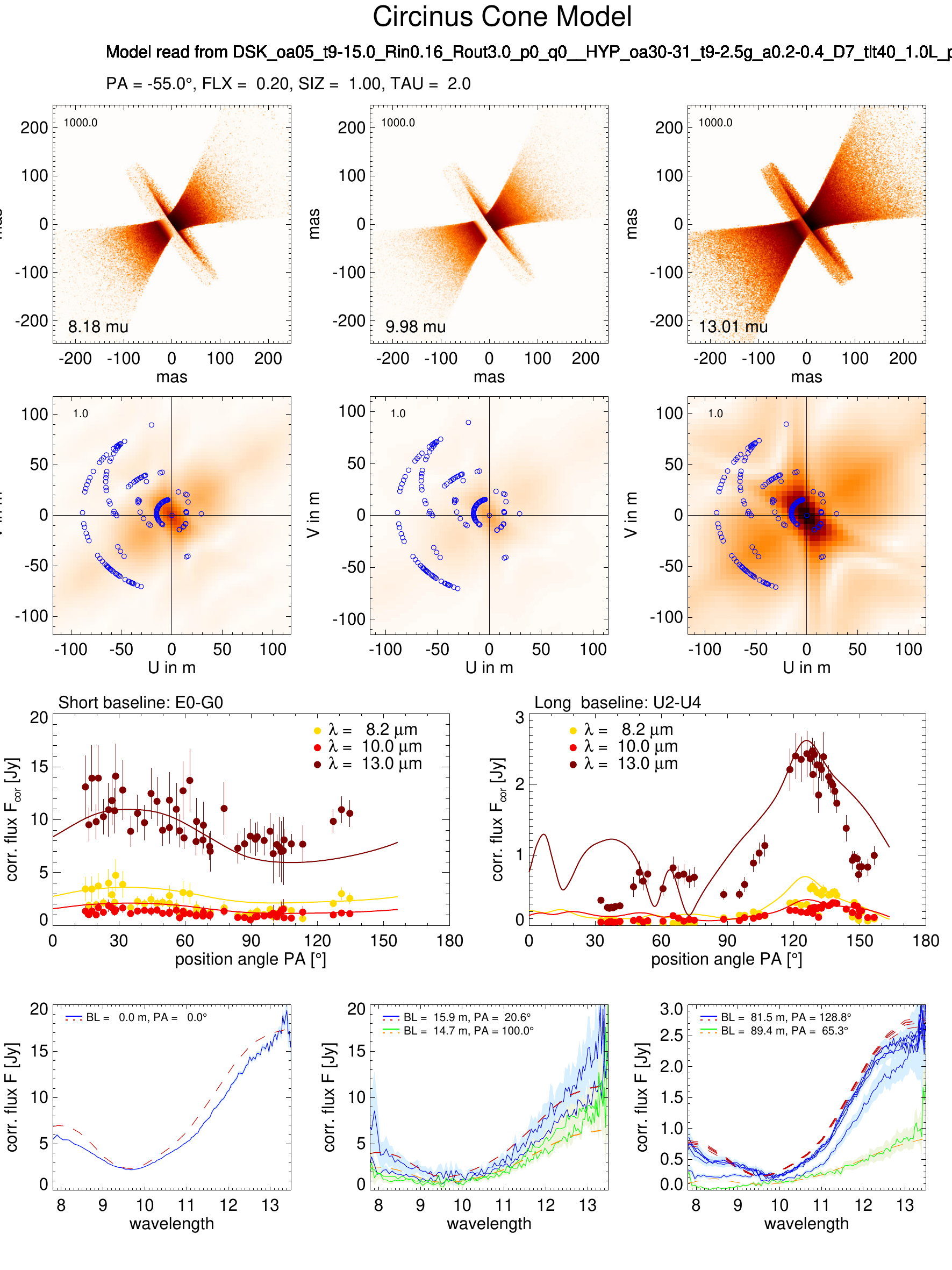}
\caption{The same as in Fig.~\ref{fig:hyp}, but for the outer hyperboloid wall at $a_{\text{out}}=0.4\,$pc.}
\label{fig:a0.4}
\end{figure*}
\begin{figure*}
\centering
\includegraphics[trim={0 0.7cm 0 1.9cm},clip,width=1.0\textwidth]{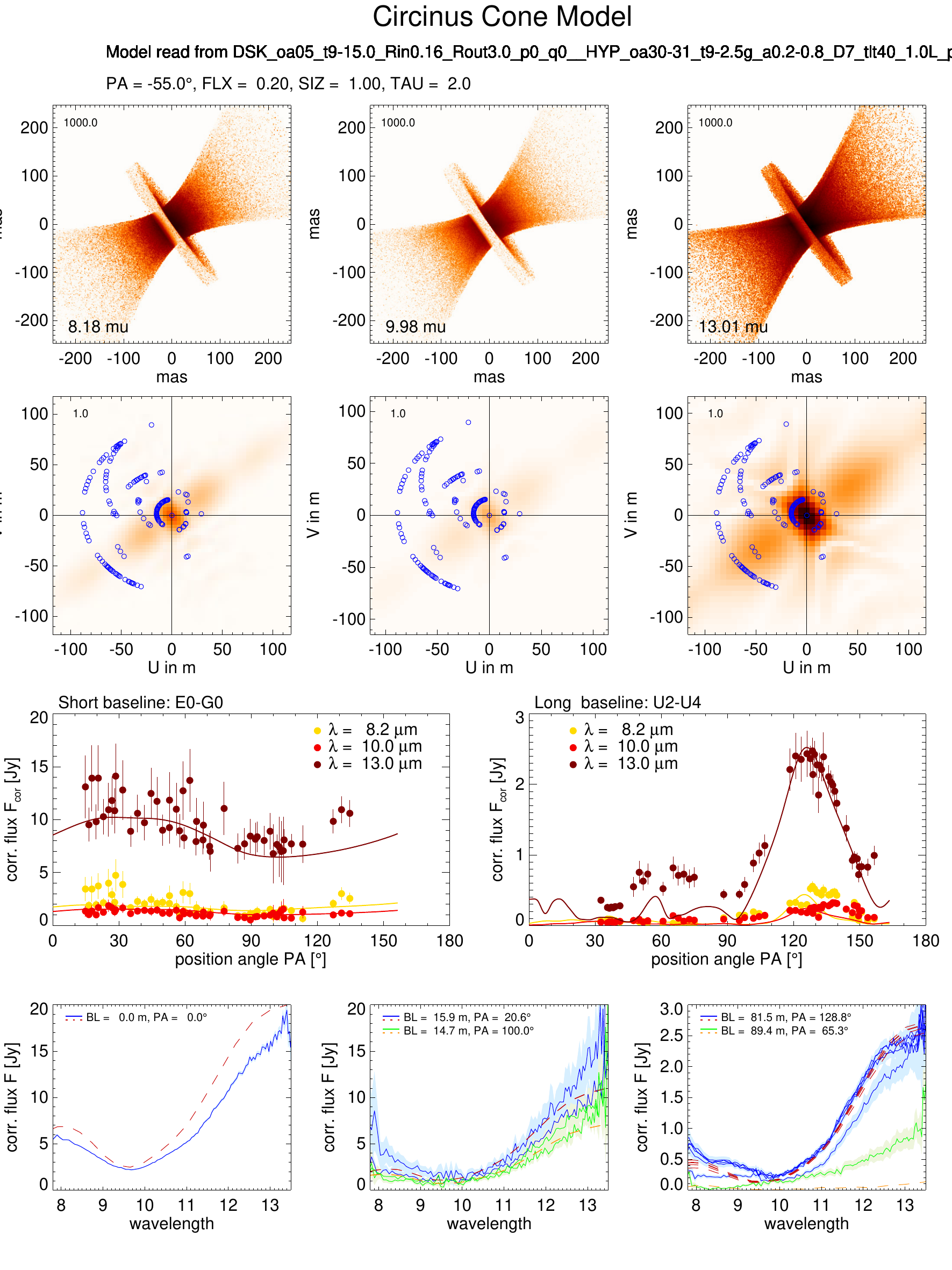}
\caption{The same as in Fig.~\ref{fig:hyp}, but for the outer hyperboloid wall at $a_{\text{out}}=0.8\,$pc.}
\label{fig:a0.8}
\end{figure*}
\begin{figure*}
\centering
\includegraphics[trim={0 0.7cm 0 1.9cm},clip,width=1.0\textwidth]{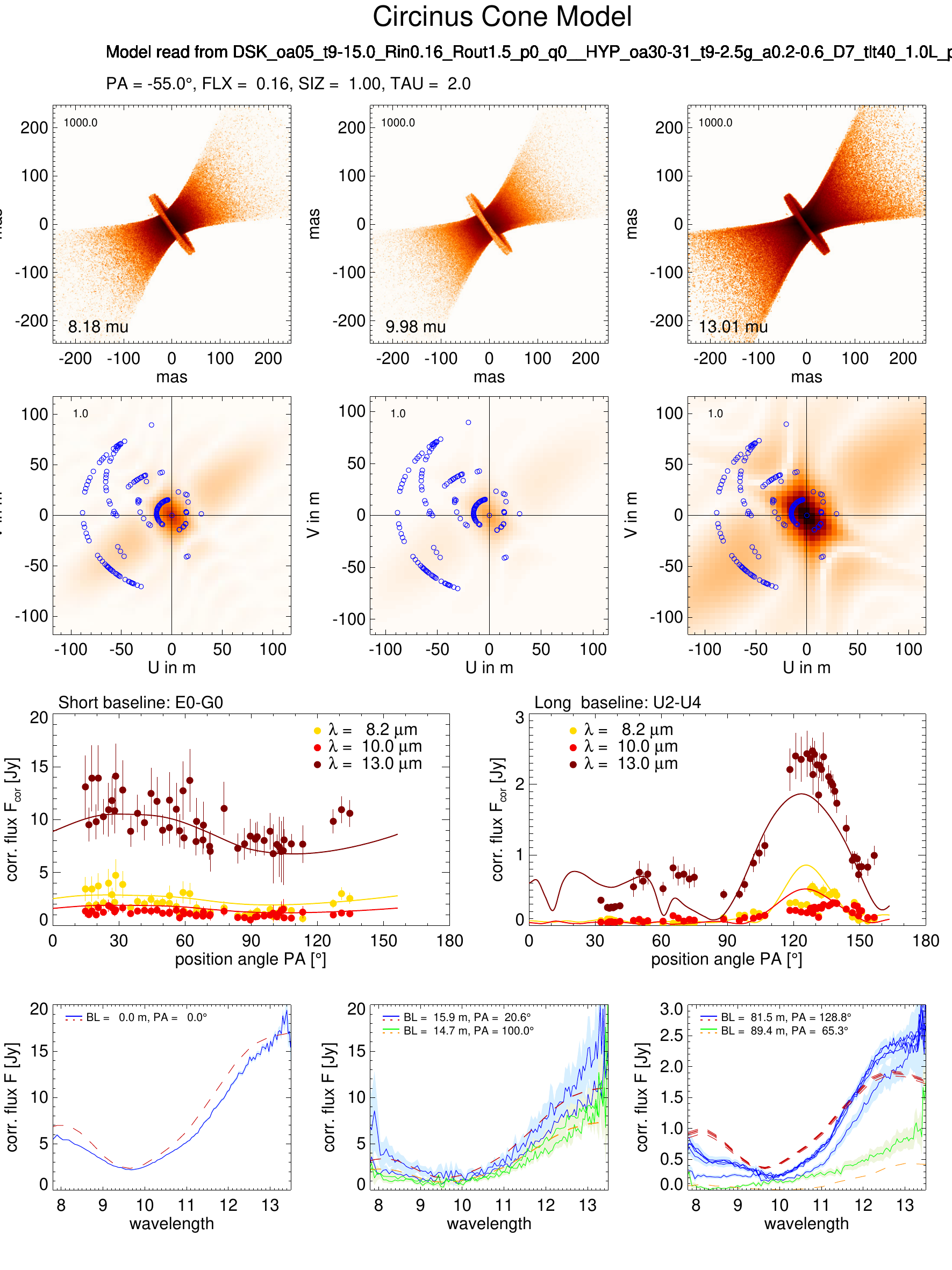}
\caption{The same as in Fig.~\ref{fig:hyp}, but for disc the outer radius $R_{\text{out}}=1.5\,$pc.}
\label{fig:R1.5}
\end{figure*}
\begin{figure*}
\centering
\includegraphics[trim={0 0.7cm 0 1.9cm},clip,width=1.0\textwidth]{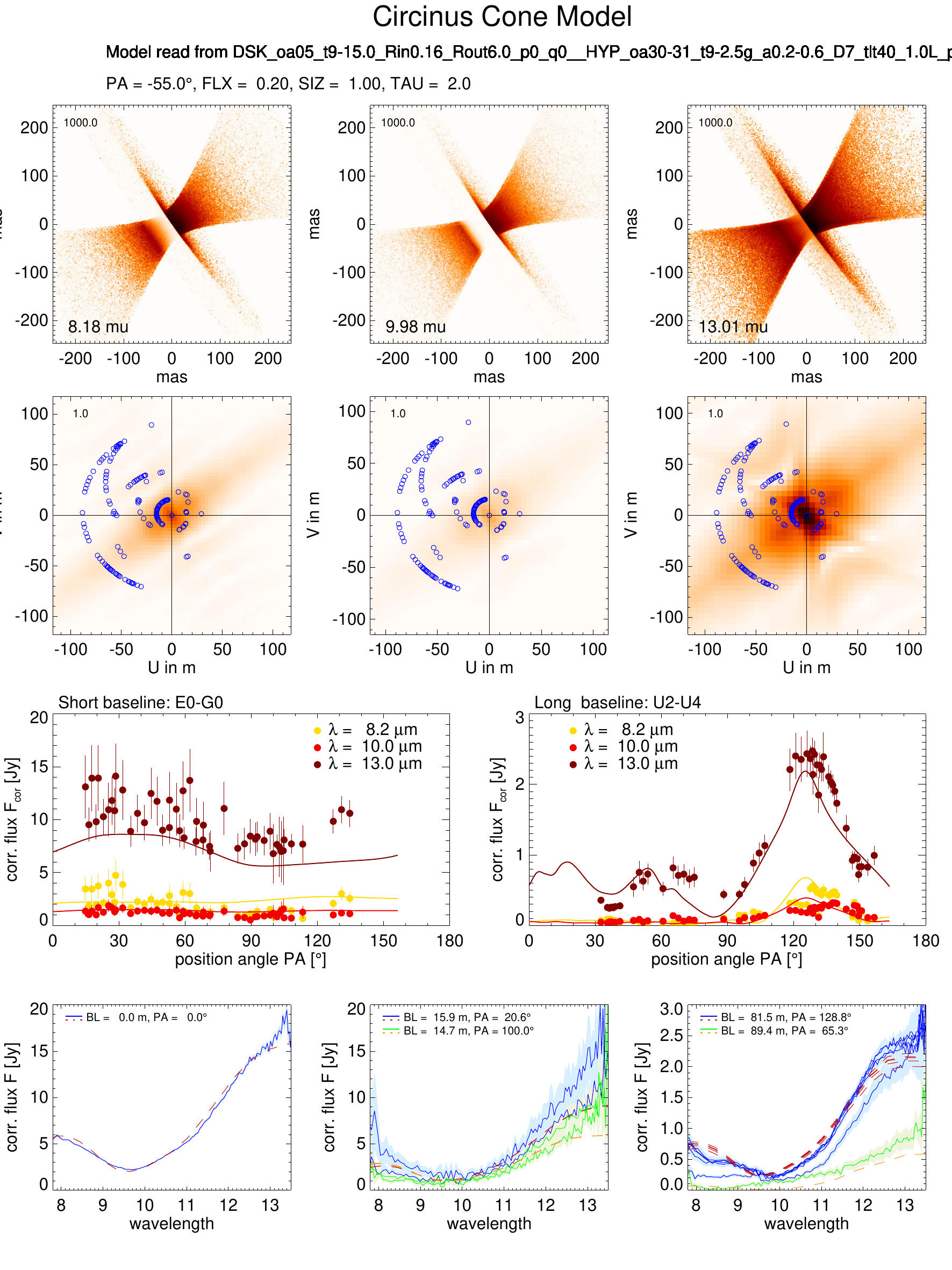}
\caption{The same as in Fig.~\ref{fig:hyp}, but for disc the outer radius $R_{\text{out}}=6\,$pc.}
\label{fig:R6}
\end{figure*}

\bsp

\label{lastpage}

\end{document}